\documentstyle[11pt,aaspp4,eqsecnum,epsf]{article}

\def\dpsi{\mathop{{d\psi\over{d\xi}}}}
\def\d2psi{\mathop{{d^2\psi\over{d\xi^2}}}}
\def\dpdr
{\mathop{\left({{\partial P_s}\over{\partial R}}\right)_{M,\sigma_c}}}
\def\dpdxi
{\mathop{\left({{\partial P_s}\over{\partial \xi}}\right)_{M,\sigma_c}}}
\def\dxidr
{\mathop{\left({{\partial \xi}\over{\partial R}}\right)_{M,\sigma_c}}}

\begin{document}

\title{A MODEL FOR THE INTERNAL STRUCTURE OF MOLECULAR CLOUD CORES}
\author{Dean E. McLaughlin and Ralph E. Pudritz}
\affil{Department of Physics and Astronomy, McMaster University \\
Hamilton, Ontario L8S 4M1, Canada \\ dean@physics.mcmaster.ca,
pudritz@physics.mcmaster.ca}
\lefthead{McLaughlin \& Pudritz}
\righthead{Internal Structure of Molecular Cores}

\begin{abstract}

We generalize the classic Bonnor-Ebert stability analysis of
pressure-truncated, self-gravitating gas spheres, to include clouds with
arbitrary equations of state. A virial-theorem analysis is also used to
incorporate mean magnetic fields into such structures.  The results are
applied to giant molecular clouds (GMCs), and to individual dense cores,
with an eye to accounting for recent observations of the internal
velocity-dispersion profiles of the cores in particular. We argue that GMCs
and massive cores are at or near their critical mass, and that in such a
case the size-linewidth and mass-radius relations between them are only
weakly dependent on their internal structures; any gas equation of state
leads to essentially the same relations. We briefly consider the possibility
that molecular clouds can be described by polytropic pressure-density
relations (of either positive or negative index), but show that these are
inconsistent with the apparent gravitational virial equilibrium, $2\ {\cal
U} +{\cal W}\approx0$, of GMCs and of massive cores. This class of models
would include clouds whose nonthermal support comes entirely from Alfv\'en
wave pressure. The simplest model consistent with all the salient features
of GMCs and cores is a ``pure logotrope,'' in which $P/P_c=1+A\ {\hbox{ln}}
(\rho/\rho_c)$.  Detailed comparisons with data are made to estimate the
value of $A$, and an excellent fit to the observed dependence of velocity
dispersion on radius in cores is obtained with $A\simeq0.2$.

\ 

{\large{\it To appear in the Astrophysical Journal}}

\end{abstract}


\section{Introduction}

Giant molecular clouds (GMCs; $M\sim10^5-10^6M_\odot$) in the Galaxy are
highly inhomogeneous: they are often filamentary in appearance, consisting
of discrete clumps, or cores ($M\la10^3M_\odot$), which contain most of the
mass of a cloud (including any young stars) and are surrounded by a more
diffuse component of predominantly atomic gas (e.g.,
\markcite{wil95}Williams, Blitz, \& Stark 1995). Indeed, GMCs are clumpy on
all scales observed, and are possibly even fractal in nature
(\markcite{fal91}Falgarone, Phillips, \& Walker 1991). Remarkably enough,
however, the gross properties of cloud complexes are rather simply
interrelated. Total masses, mean densities, and average velocity dispersions
vary with sizes (effective radii) roughly as $M\propto R^2$, $\rho_{\rm{ave}}
\propto R^{-1}$, and $\sigma_{\rm{ave}}\propto R^{1/2}$
(\markcite{lar81}Larson 1981; \markcite{san85}Sanders, Scoville, \& Solomon
1985; \markcite{sol87}Solomon et al.~1987), with uncertainties in the
exponents typically of order $\pm$0.1. The seeming universality of these
results demands a physical explanation.

The relationship between size and linewidth (which term we use
interchangeably with velocity dispersion) is further interesting because
$\sigma$ is observed to decrease towards smaller radii {\it inside} GMCs,
and within individual dense cores (\markcite{lar81}Larson 1981;
\markcite{mie94}Miesch \& Bally 1994; \markcite{ful92}Fuller \& Myers 1992;
\markcite{cas95}Caselli \& Myers 1995). Since the total linewidths of GMCs
are mostly nonthermal (the clouds are stable against gravitational collapse
on the largest scales, but have masses several orders of magnitude above the
thermal Jeans value, so that their support must come largely from nonmagnetic
or, very likely, MHD turbulence), this decrease reflects a move towards
domination by thermal motions on the smallest scales. We should therefore
expect the linewidths of low-mass cores to have a smaller turbulent
component than those of high-mass cores. This is indeed the case; in fact,
the velocities nearest the centers of small cores are almost (to within a
few percent) wholly thermal. However, it also happens
(\markcite{cas95}Caselli \& Myers 1995) that the nonthermal velocity
dispersion shows {\it a stronger dependence on radius in low-mass cores}
(where $\sigma_{\rm{NT}}\propto r^{0.5}$) than in massive ones
($\sigma_{\rm{NT}}\propto r^{0.2}$). Because star formation is localized in
the cores of GMCs, their overall structure --- and this aspect specifically
--- bears strongly on our understanding of this most fundamental process.

The goal of this paper is to find a model for the internal structure of
molecular cores (low- and high-mass both) which quantitatively matches their
observed, internal velocity-dispersion profiles, and is also consistent with
global properties (such as the observed mass-radius-linewidth relations) of
large, self-gravitating clumps and even whole GMCs. Our approach is to
reduce this problem to the specification of a (total) pressure-density
relation --- an equation of state --- that, when used to solve the equation
of hydrostatic equilibrium, results in a gas cloud with the required
features. It is significant that the linewidth profiles of cores are
insensitive to the presence or absence of young stars, and thus may be
viewed as one of the prerequisites for star formation (\markcite{ful92}Fuller
\& Myers 1992; \markcite{cas95}Caselli \& Myers 1995). This justifies our
focus on the structure of purely gaseous clouds.

To proceed, we shall resort to a virial-theorem treatment of molecular clouds
which idealizes them as spheres of gas in hydrostatic equilibrium and
satisfying Poisson's equation. This is appropriate enough for dense cores,
which in many cases are roughly spheroidal (probably prolate:
\markcite{mye91}Myers et al.~1991), or even near spherical (e.g.,
\markcite{wil95}Williams et al.~1995), overall. (Although their internal
density distributions may not be especially smooth on very small scales, we
concern ourselves here with a description of their {\it bulk} structure.) In
addition, observations of very massive cores imply that they are
self-gravitating and in approximate ``gravitational virial equilibrium,''
$2\ {\cal U}+{\cal W}\approx0$. This is not the case for low-mass cores, but
these still appear to satisfy the full virial theorem if surface-pressure
terms are included (\markcite{ber92}Bertoldi \& McKee 1992). Finally, even
the total masses, radii, and linewidths of entire GMC complexes are generally
consistent with virial-equilibrium models of spheres (e.g.,
\markcite{sol87}Solomon et al.~1987; \markcite{elm89}Elmegreen 1989); and it
has been repeatedly confirmed that whole clouds tend to comply with $2\ {\cal
U}+{\cal W}\approx0$ (\markcite{lar81}Larson 1981; \markcite{sol87}Solomon et
al.~1987; \markcite{my88b}Myers \& Goodman 1988b).

It should be noted that the correct, average mass-radius and size-linewidth
relations can be recovered in a purely scale-free description of GMCs (e.g.,
\markcite{hen91}Henriksen 1991). But the distinctly different scalings of
velocity dispersion with radius inside low- and high-mass clumps is one
indication that molecular clouds are in fact not entirely featureless. Other
current models for these objects (such as isothermal spheres or
negative-index polytropes) face similar difficulties in accounting
simultaneously for their global properties and their internal structures.

Magnetic fields are an important presence in regions of star formation
(e.g., \markcite{hei93}Heiles et al.~1993), so we begin in \S2 by writing
down the virial-theorem (or Bonnor-Ebert) relations between the masses and
total linewidths --- including turbulent velocities --- of magnetized
spheres, truncated at radii such that the internal pressure just balances
that of a diffuse surrounding medium. We then develop a stability
criterion for such clouds which depends only on the assumption that their
thermal linewidths (i.e., kinetic temperatures) are invariant. When combined
with the circumstantial evidence for equipartition between the kinetic and
mean-field magnetic energies in GMCs and massive cores (\markcite{my88a}Myers
\& Goodman 1988a, \markcite{my88b}b; \markcite{ber92}Bertoldi \& McKee
1992), this stability criterion leads to mass-radius-linewidth relations
between critical-mass objects that agree with the observed scalings (both the
exponents and the coefficients) among GMCs. Our analysis is therefore similar
to those of, e.g., \markcite{chi87}Chi\`eze (1987), \markcite{fle88}Fleck
(1988), and especially \markcite{elm89}Elmegreen (1989), but ours holds for
clouds with an {\it arbitrary} equation of state. In addition, we find that
critically stable clouds in magnetic equipartition should all satisfy $2\ 
{\cal U} +{\cal W}\approx0$. We therefore conclude that GMCs and massive
cores are approximately at their critical masses and magnetically
supercritical, with $M_{\rm{crit}}\simeq 2\ M_{\Phi}$ (see also
\markcite{mck89}McKee 1989; \markcite{ber92}Bertoldi \& McKee 1992). The
generality of these results allows for the investigation of essentially any
pressure-density relation as a potential description of the interiors of
molecular cores, just so long as critically stable configurations are at all
possible.

In \S3, we discuss polytropic equations of state: $P\propto\rho^{\gamma}$,
with $\gamma$ any positive number. Clearly, if $\gamma\geq1$, then the total
velocity dispersion ($\sigma^2=P/\rho$) stays constant or decreases with
decreasing density (increasing radius) inside a cloud. If instead $\gamma$
is allowed to be less than 1 (e.g., \markcite{mal88}Maloney 1988), then
$\sigma^2$ increases with radius, as required. However, in our analysis, any
such ``negative-index'' polytrope would be unconditionally stable against
gravitational collapse; or equivalently (but independently of any stability
criterion), it could not self-consistently satisfy the relation $2\ {\cal U}
+{\cal W}\approx0$. While this is not a problem for low-mass cores, it is
inconsistent with what we know of very large clumps, and GMCs overall. Thus,
we argue that polytropic pressure-density relations give an incomplete
picture of interstellar clouds. A corollary to this is that weakly damped
Alfv\'en waves, for which $P\propto\rho^{1/2}$ (\markcite{mck95}McKee \&
Zweibel 1995), cannot be invoked as the {\it sole} explanation of nonthermal
linewidths in GMCs.

Section 4 describes what is, in our view, a more suitable alternative. There
we consider the possibility that pressure varies only logarithmically with
density: $P/P_c=1+A\ {\hbox{ln}}(\rho/\rho_c)$. We refer to the resulting
gas cloud as a ``pure'' logotrope. This term was introduced by
\markcite{liz89}Lizano \& Shu (1989; see also \markcite{mck89}McKee 1989),
who actually added a logarithmic term to an otherwise isothermal equation of
state in an attempt to account for turbulent linewidths. Such models have
also been studied in detail by \markcite{geh96}Gehman et al.~(1996).  Here,
however, we dispense with the explicitly isothermal component, for two
reasons: (1) Assuming $P=\rho\sigma_{\rm{T}}^2+\kappa\ {\hbox{ln}}
(\rho/\rho_{\rm{ref}})$, with $\sigma_{\rm{T}}$ the thermal velocity
dispersion and $\rho_{\rm{ref}}$ some reference density
(\markcite{liz89}Lizano \& Shu 1989), the observational inference that
linewidths should be essentially purely thermal at the centers of cores
requires that $\rho_{\rm{ref}}=\rho_c$. But then the internal
$\sigma^2=P/\rho$ {\it decreases} with increasing radius. (2) If now
$P/P_c=\rho/\rho_c+\kappa\ {\hbox{ln}}(\rho/\rho_c)$, then for large values
of $\kappa$ such as those suggested by \markcite{geh96}Gehman et al.~(1996),
$P$ vanishes for $\rho/\rho_c$ rather near unity. Thus, real clouds would
have to be almost uniform-density, which again is not observed.

These difficulties do not extend to the specific equation of state which we
examine.  Instead, an outwards-increasing velocity-dispersion profile
obtains for an equilibrium pure logotrope. Such a model can moreover account
for the observed mass-radius and size-linewidth relations between GMCs.  We
also demonstrate that the linewidth measurements in both low- and high-mass
cores, from a variety of molecular clouds, are quantitatively reproduced if
$A\simeq0.2$. The argument makes explicit use of the fact that small cores
are {\it not} at their critical masses (while large ones generally are), but
that they are still in approximate virial equilibrium when the effects of
surface pressures are considered. 

Although the logarithmic $P-\rho$ relation we advocate is phenomenological,
its overall viability, along with the failings of other models, ultimately
makes for a useful description of molecular clouds and their cores.

\section{Generalized Bonnor-Ebert Relations}

Interstellar clouds can be viewed as essentially ``pressure-truncated''
bodies of gas. (This term is meant to imply the existence, not of some radius
where a gas cloud suddenly ends, but of one where it ``blends in'' with an
ambient medium.) Thus, the boundary of a GMC is set by pressure balance with
the surrounding, hot ISM. Note that even if there is no such balance
initially, it will eventually obtain after an overall expansion or
contraction of the cloud complex. The extent of a core within a GMC is
similarly limited by the pressure of a tenuous interclump medium, as has
been argued by \markcite{ber92}Bertoldi \& McKee (1992). The equilibrium
structure of pressure-truncated isothermal spheres was first described by
\markcite{ebe55}Ebert (1955) and \markcite{bon56}Bonnor (1956) (see also
\markcite{mcc57}McCrea 1957). Their expressions connecting the masses,
radii, and linewidths of the spheres follow from the assumption of
hydrostatic equilibrium and Poisson's equation, and therefore are written in
terms of an internal pressure profile and gravitational potential. In
Appendix A, we re-derive these Bonnor-Ebert relations, but for clouds
satisfying an arbitrary gas equation of state
(eqs.~[\ref{eq:4a}]--[\ref{eq:4d}]). We also provide a connection
(eqs.~[\ref{eq:13a}]--[\ref{eq:13}]) with the more transparent and
observationally convenient virial-theorem formulation given by equations
(\ref{eq:12}) below. As various authors have noted (\markcite{chi87}Chi\`eze
1987; \markcite{fle88}Fleck 1988; \markcite{mal88}Maloney 1988;
\markcite{elm89}Elmegreen 1989), these relations in either guise provide a
framework for an understanding of the standard ``Larson's laws'' for GMCs
($M\propto R^2$ and $\sigma\propto R^{1/2}$; \markcite{lar81}Larson 1981).

The development of Appendix A refers specifically to ``nonmagnetic'' clouds,
meaning only that no {\it ordered} (mean) magnetic field is considered to be
present. We work on the assumption that the effects of disordered fields (MHD
turbulence) can be separately dealt with, in an equivalent hydrostatic
problem that makes use of an effective equation of state to describe
{\it all} of the contributions to gas pressure as a function of density.
Still, the effects of mean fields must also be considered in any applications
to real interstellar clouds.

\subsection{Magnetic Equilibria}

In the absence of any analytic models for magnetohydrostatic clouds, we
proceed by assuming spherical symmetry and turning to the scalar virial
theorem:
\begin{equation}
2\ {\cal U}(1-P_s/P_{\rm{ave}}) + {\cal M} + {\cal W}\ =\ 0\ .
\label{eq:5}
\end{equation}
Here the mass-averaged, total one-dimensional velocity dispersion of a cloud
with radius $R$ is related to its mean pressure and density by
\begin{equation}
\sigma_{\rm{ave}}^2={{\int_0^{R}4\pi r^2\rho\sigma^2\ dr} \over{\int_0^{R}
4\pi r^2\rho\ dr}}={P_{\rm{ave}}\over{\rho_{\rm{ave}}}}\ ,
\label{eq:6}
\end{equation}
so that the kinetic (or internal), mean-field magnetic, and gravitational
energies are given by the usual
\begin{equation}
{\cal U}={3\over{2}}\ M\sigma_{\rm{ave}}^2\ ,
\label{eq:7}
\end{equation}
\begin{equation}
{\cal M}={1\over{8\pi}}\int B^2\ dV + {1\over{4\pi}}\oint ({\bf r}\cdot
{\bf B}){\bf B}\cdot d{\bf S} - {1\over{8\pi}}\oint B^2{\bf r}\cdot d{\bf S}
\ ,
\label{eq:8}
\end{equation}
and
\begin{equation}
{\cal W}=-G\int_0^R{{m\ dm}\over{r}}\equiv-{3\over{5}}a{{GM^2}\over{R}}\ .
\label{eq:9}
\end{equation}
The parameter $a$ is essentially a measure of the non-uniformity of a gas
sphere, and as such depends on the equation of state and the truncation
radius (in terms of a fixed scale $r_0$; eq.~[\ref{eq:13a}]). However, it is
generally of order unity: for a power-law density profile $\rho\propto
r^{-p}$, equation (\ref{eq:9}) gives $a=(1-p/3)/(1-2p/5)$. We expect $1\la
p\la2$ in a realistic GMC or core, and thus $10/9\la a\la5/3$.

It is often more useful to work in terms of the virial parameter of
\markcite{ber92}Bertoldi \& McKee (1992):
\begin{equation}
\alpha_{\rm{mag}}\equiv{{5\sigma_{\rm{ave}}^2R}\over{GM}}\ .
\label{eq:10}
\end{equation}
This {\it observable} quantity can also be written (cf.~\markcite{mck92}McKee
\& Zweibel 1992) as
$$\alpha_{\rm{mag}}=2a\ {{\cal U}\over{|{\cal W}|}}={a\over{1-P_s/
P_{\rm{ave}}}}\left(1-{{\cal M}\over{|{\cal W}|}}\right)\ .$$
Although this form of the virial theorem is appropriate for clouds of any
shape, the corrective factors for spheroidal clouds are rather near
unity (\markcite{ber92}Bertoldi \& McKee 1992), and we still allow only for
spherical symmetry here. It is further convenient to distinguish between the
virial parameters which would obtain for a cloud with and without a mean
magnetic field; these are related by
\begin{equation}
\alpha_{\rm{mag}}=\alpha_{\rm{non}}\left(1-{{\cal M}\over{|{\cal W}|}}\right)
\ .
\label{eq:11}
\end{equation}
Again, $\alpha_{\rm{mag}}$ refers specifically to the combination
(\ref{eq:10}) of observables. On the other hand, $\alpha_{\rm{non}}$ applies
to the ${\cal M}=0$ (no mean field) counterpart of a given cloud; it is
directly observable only in this special case (since then $\alpha_{\rm{mag}}=
\alpha_{\rm{non}}$), but may always be calculated for a given gas equation of
state, as outlined in Appendix A. Once this is known, $\alpha_{\rm{mag}}$
follows with the specification of a mean-field configuration (see \S2.3).

The connection (\ref{eq:11}) between $\alpha_{\rm{mag}}$ and
$\alpha_{\rm{non}}$ is valid insofar as the ratio $P_s/P_{\rm{ave}}$ does not
change drastically upon the ``addition'' of a mean magnetic field to a cloud
which is already in hydrostatic equilibrium. This must hold along field lines
anyway (force balance is required in that direction), and thus everywhere on
the surface of a roughly spherical cloud. Defining $a_{\rm{eff}}\equiv a(1-
{\cal M}/|{\cal W}|)$, such reasoning implies that $(\alpha_{\rm{mag}}-
a_{\rm{eff}})/\alpha_{\rm{mag}}=(\alpha_{\rm{non}}-a)/\alpha_{\rm{non}}$, so
manipulation of the virial theorem gives the following:
\begin{equation}
M=25\sqrt{{3\over{20\pi}}}\left({{\alpha_{\rm{non}}-a}\over
{\alpha_{\rm{non}}}}\right)^{1/2}{1\over{\alpha_{\rm{mag}}^{3/2}}}\ 
{{\sigma_{\rm{ave}}^4}\over{(G^3P_s)^{1/2}}}\ ,
\label{eq:12}
\eqnum{\ref{eq:12}a}
\end{equation}
\begin{equation}
R=5\sqrt{{3\over{20\pi}}}\left({{\alpha_{\rm{non}}-a}
\over{\alpha_{\rm{non}}}}\right)^{1/2}{1\over{\alpha_{\rm{mag}}^{1/2}}}\ 
{{\sigma_{\rm{ave}}^2}\over{(GP_s)^{1/2}}}\ ,
\eqnum{\ref{eq:12}b}
\end{equation}
\begin{equation}
\Sigma=\sqrt{20\over{3\pi}}\ \left({{\alpha_{\rm{non}}}
\over{\alpha_{\rm{non}}-a}}\right)^{1/2}{1\over{\alpha_{\rm{mag}}^{1/2}}}\
\left({P_s\over{G}}\right)^{1/2}\ ,
\eqnum{\ref{eq:12}c}
\end{equation}
and
\begin{equation}
\rho_{\rm{ave}}={{\alpha_{\rm{non}}}\over{\alpha_{\rm{non}}-a}}\ {P_s\over
{\sigma_{\rm{ave}}^2}}\ .
\eqnum{\ref{eq:12}d}
\addtocounter{equation}{+1}
\end{equation}
These relations, of which only two are independent (they are contained, for
example, in eqs.~[9] and [10] of \markcite{elm89}Elmegreen 1989), are also
given by \markcite{har94}Harris \& Pudritz (1994) for the specific case of
critically stable isothermal spheres (as defined in \S2.2; $\alpha_{\rm{non}}
=2.054$ and $a=1.221$ at a cloud radius $R_{\rm{crit}}/r_0=2.150$). As
written here, they apply to any generic gas cloud, stable or unstable,
isothermal or not.

Any cloud in hydrostatic equilibrium and satisfying Poisson's equation has a
central region where the potential, density, and velocity dispersion are
very nearly constant with radius. Equations (\ref{eq:13a}) and (\ref{eq:13b})
show that spheres truncated in or just outside this central region (i.e., at
radii small enough that $P_s\approx P_c$) have $\alpha_{\rm{non}}\gg a$, and
thus $\alpha_{\rm{mag}}\propto M^{-2/3}\propto R^{-2}$ by equations
(\ref{eq:12}a, b). Roughly this scaling has been observed for the virial
parameters of low-mass, high-$\alpha$ cores in several GMCs, leading
\markcite{ber92}Bertoldi \& McKee (1992) to argue that such clumps can be
viewed as truncated spheroids which are essentially in pressure equilibrium
with an intracloud medium.

\subsection{Stability: Critical Clouds}

Given equations (\ref{eq:12}) for the equilibrium structure of
pressure-truncated gas spheres, we are in a position to question their
stability:  Under what conditions will they be able to withstand
the combined effects of self-gravity and surface pressure, and when will they
be unstable to wholesale gravitational collapse? The answer to this depends,
of course, on any boundary conditions attached to a perturbation of the
cloud. Obviously, the total mass should be unchanged by a contraction or
expansion of the entire structure. In addition, following
\markcite{mal88}Maloney (1988), we suppose that the central velocity
dispersion remains constant as the cloud radius, or the surface pressure, is
varied. This stipulation is meant to reflect the fact that the turbulent
linewidth decreases steadily towards smaller scales in cores and in entire
GMCs. We therefore identify the central velocity dispersion with the thermal
part of the total linewidth: $\sigma_c^2=kT/\mu m_H$. Insisting that this be
invariant amounts to recognizing the rough uniformity of kinetic
temperatures $T\sim10K$ (which to first order can be understood as a
consequence of the competition between cosmic-ray heating and CO cooling)
over a large range of scales in interstellar clouds. The stability criterion
that follows ultimately leads to a set of results which self-consistently
explain some important observational features of GMCs and massive cores.

A gas cloud will be stable against radial perturbations if the derivative
$\partial P_s/\partial R$ (taken with $\sigma_c$ and $M$ held fixed) is
$\leq0$: a slight decrease in the cloud radius then leads to an increase in
the pressure just inside its boundary, which in turn leads to reexpansion.
Appendix B shows that, for any equation of state, this condition is just
\begin{equation}
\dpdr=-6{{P_s}\over{R}}\left[{{1-(5/6)(\alpha_{\rm{non}}-a)^{-1}}
\over{3-\rho_{\rm{ave}}/\rho_s}}\right]\leq0\ ,
\label{eq:14}
\end{equation}
where $\rho_s$ is the internal density at the edge of the cloud. Although
this stability criterion has been derived without explicitly considering the
effects of mean magnetic fields, we expect that it should not be greatly
altered by their inclusion. (This is again implied by our assumption of
approximate spherical symmetry, since eq.~[\ref{eq:14}] must at least be
satisfied along field lines at the boundary of a magnetized cloud.)

Depending on the equation of state, there may exist a radius for which a
pressure-truncated cloud is marginally stable (the expression [\ref{eq:14}]
is just 0), and beyond which it is unstable. It is these critical
equilibria, which must have
\begin{equation}
\alpha_{\rm{non}}-a={5\over{6}}\ ,
\label{eq:15}
\end{equation}
that are of particular interest here. Once $\alpha_{\rm{non}}$ and $a$ are
known as functions of radius (as in Appendix A), the satisfaction, if
possible, of equation (\ref{eq:15}) sets the boundary $R_{\rm{crit}}/r_0$ of
the cloud. This in turn allows for evaluation of the coefficients in
equations (\ref{eq:12}). Quite generally,
\begin{equation}
M_{\rm{crit}}={{25}\over{\sqrt{8\pi}}}\ 
{1\over{\alpha_{\rm{non}}^{1/2}\alpha_{\rm{mag}}^{3/2}}}\ 
{{\sigma_{\rm{ave}}^4}\over{(G^3P_s)^{1/2}}}\ ,
\label{eq:16}
\eqnum{\ref{eq:16}a}
\end{equation}
\begin{equation}
R_{\rm{crit}}={5\over{\sqrt{8\pi}}}{1\over{\alpha_{\rm{non}}^{1/2}
\alpha_{\rm{mag}}^{1/2}}}\ {{\sigma_{\rm{ave}}^2}\over{(GP_s)^{1/2}}}\ ,
\eqnum{\ref{eq:16}b}
\end{equation}
\begin{equation}
\Sigma_{\rm{crit}}=\sqrt{{8\over{\pi}}}\ \left({{\alpha_{\rm{non}}}
\over{\alpha_{\rm{mag}}}}\right)^{1/2}\left({{P_s}\over{G}}\right)^{1/2}\ ,
\eqnum{\ref{eq:16}c}
\end{equation}
and
\begin{equation}
\rho_{\rm{ave,crit}}={{6\alpha_{\rm{non}}}\over{5}}\ {{P_s}
\over{\sigma_{\rm{ave}}^2}}\ .
\eqnum{\ref{eq:16}d}
\addtocounter{equation}{+1}
\end{equation}
Equation (\ref{eq:16}c) shows that a nonmagnetic cloud ($\alpha_{\rm{mag}}=
\alpha_{\rm{non}}$) on the verge of gravitational collapse has a mean column
density which is fixed by the pressure of the surrounding medium,
independently of any gas equation of state.

In general, the virial parameter of a given cloud is sensitive to its
internal structure (through the equation of state) and its total radius.
However, for a {\it critically stable} cloud we always have
$\alpha_{\rm{non}}=a+5/6$. Since $a$ is typically of order (but slightly
greater than) unity, this implies $\alpha_{\rm{non}}\approx2$ and $1-P_s/
P_{\rm{ave}}=a/\alpha_{\rm{non}}\approx1/2$. The virial theorem (\ref{eq:5})
then becomes
\begin{equation}
{\cal U}+{\cal M}+{\cal W}\approx0\ .
\label{eq:17a}
\end{equation}
If there is equipartition ${\cal M}\approx{\cal U}$ between the magnetic and
kinematic energies in such a cloud, then $2\ {\cal U}+{\cal W}\approx0$ as
well; thus, ${\cal M}/|{\cal W}|\approx1/2$, and the critical
$\alpha_{\rm{mag}}$ is expected to be of order unity (cf.~eq.~[\ref{eq:11}];
see also \markcite{elm89}Elmegreen 1989 and \markcite{mck92}McKee \& Zweibel
1992).

Observations of molecular clouds show that $2\ {\cal U}+{\cal W}\approx0$ (or
equivalently, $\alpha_{\rm{mag}}\approx1$), and are {\it consistent} with
${\cal M}\approx{\cal U}$ (\markcite{my88a}Myers \& Goodman 1988a,
\markcite{my88b}b), although actual magnetic-field measurements are few and
uncertain. The most massive cores in GMCs similarly tend to show $2\ {\cal U}
+{\cal W}\approx0$ and $\alpha_{\rm{mag}}$ near 1 (this is not the case for
low-mass cores, however:  e.g., \markcite{wil94}Williams, de Geus, \& Blitz
1994; \markcite{wil95}Williams et al.~1995; see also \S4.1 below).
Observations of them are also indicative of magnetic equipartition
(\markcite{ber92}Bertoldi \& McKee 1992), though again the evidence is
rather indirect, and not necessarily conclusive. Having said this, it does
seem that GMCs and massive cores both satisfy equation (\ref{eq:17a}), which
would imply that they are at or near their critical masses.

Thus arrived at, this conclusion depends on the criterion one adopts for
cloud stability, i.e., it follows from the assumption that the thermal
linewidth $\sigma_c$ is held fixed during any radial perturbation (and from
the additional proviso that ${\cal M}\approx{\cal U}$). Nevertheless, our
analysis --- which specifies no equation of state --- provides a natural
explanation for the fact that so many GMCs and large cores appear to be in
simple ``gravitational virial equilibrium'' (i.e., $2\ {\cal U}+{\cal W}
\approx0$), even though they are threaded by appreciably strong mean
magnetic fields.

There is also separate evidence for the criticality of massive molecular
cores. For instance, \markcite{ber92}Bertoldi \& McKee (1992), in their
study of the cores in four GMCs, argue that the most massive are at least
magnetically supercritical, a necessary condition for gravitational
instability.  More fundamentally, star-forming regions clearly must be
susceptible to gravitational collapse; but in the Rosette GMC at least,
those cores which are most obviously associated with IRAS sources are also
among the heaviest (\markcite{wil95}Williams et al.~1995). As mentioned
above, the largest cores in several GMCs also have the smallest virial
parameters (as low as 1), so that this would seem to be a feature of clouds
which are close to instability. The observation of a mean $\alpha_{\rm{mag}}
\approx1$ for GMCs then implies that they, too, are near some critical mass,
and certainly in excess of the nonmagnetic Jeans or Bonnor-Ebert value
(for which ${\cal M}=0$ implies ${\cal U}+{\cal W}\approx0$ and
$\alpha_{\rm{mag}}=\alpha_{\rm{non}}\approx2$). \markcite{mck89}McKee (1989)
has further argued that GMCs on the whole are magnetically supercritical,
and of course they must be strongly self-gravitating in order to be
molecular at all (e.g., \markcite{elm85}Elmegreen 1985).

On a related note, \markcite{mck89}McKee (1989) points out that GMCs must
generally be near criticality because they show a $P_{\rm{ave}}$ which is
typically an order of magnitude larger than the total (thermal plus
turbulent) pressure in the hot ISM. Given our stability criterion, equation
(\ref{eq:16}d) shows that self-gravity can supply a maximal pressure
enhancement $P_{\rm{ave}}/ P_s\approx2.5$ of a spherical cloud over its
surrounding medium, and this only for a critical-mass body. Even putting the
nonsphericity of GMCs aside, however, it is important to note that the
analysis here speaks only to the molecular, self-gravitating parts of GMCs,
and not to their diffuse, low-$A_V$ \ion{H}{1} components.
\markcite{elm89}Elmegreen (1989) has shown that the weight of these atomic
``envelopes'' can easily increase the pressure at the boundaries of the
molecular parts of a cloud complex by a factor of 5 or more above the value
in the ISM at large; overall, then, $P_{\rm{ave}}/ P_{\rm{ISM}}>10$.

If GMCs and their most massive cores are indeed critical-mass objects,
then they must all have the same dimensionless radii (although the physical
scale $r_0$ will generally vary), and the same virial parameters. Aside
from possible variations in $P_s$, which are discussed in detail by
\markcite{elm89}Elmegreen (1989), this causes the coefficients in equations
(\ref{eq:16}) or (\ref{eq:12}) to be roughly constant, and allows for well
defined mass-radius-linewidth relations between clouds. Moreover, we have
argued that $\alpha_{\rm{non}}\simeq2$ and, in the event of magnetic
equipartition, $\alpha_{\rm{mag}}\simeq1$ for critical clouds, regardless of
the underlying equation of state. The coefficients in the
$M-R-\sigma_{\rm{ave}}$ scalings are then independent of this detail, and
the average properties of GMCs can shed no light on their internal structure.
This both explains why critical, magnetized isothermal-sphere models are
successful in quantitatively accounting for the observed scalings (with
$\alpha_{\rm{non}}=2.054$ and $\alpha_{\rm{mag}}\simeq1$:
\markcite{elm89}Elmegreen 1989; \markcite{har94}Harris \& Pudritz 1994), and
implies that the same agreement with the data comes with {\it any} model
which provides for the existence of a critical mass (see eqs.~[\ref{eq:411}]
below).

Finally, equations (\ref{eq:16}) as written would suggest that all GMCs must
be under similar surface pressures $P_s$ if, for example, $\Sigma$ is to be
roughly the same among them. Other authors (e.g., \markcite{mck89}McKee
1989; \markcite{mou87}Mouschovias 1987; \markcite{my88b}Myers \& Goodman
1988b) have argued instead that either $\Sigma$ itself (and hence
$P_{\rm{ave}}$), or the mean field strength $B_{\rm{ave}}$, is the more
fundamentally invariant attribute of GMCs.  If so, then $P_s$ could be
eliminated from the critical Bonnor-Ebert relations in favor of any of these
quantities, and our approach does not preclude the others.

\subsection{Magnetic Field Model}

Consider a cloud of radius $R$, threaded by a mean magnetic field
approximated as uniform and of magnitude $B_{\rm{ave}}$. Outside the cloud,
let $B$ fall off as $r^{-2}$ to a radius $R_0$, where it matches onto an
ambient, uniform field of strength $B_0$ (\markcite{nak84}Nakano 1984).
Conservation of flux (or continuity of the normal component of {\bf B}
across the boundary of the cloud) demands $B_{\rm{ave}}R^2=B_0R_0^2$, and
evaluating equation (\ref{eq:8}) at the surface $r=R_0$ gives
$${\cal M}={{B_{\rm{ave}}^2R^3}\over{3}}\ \left(1-{R\over{R_0}}\right)\ .$$
Defining $\beta=8\pi P_{\rm{ave}}/B_{\rm{ave}}^2$ and $\Phi=\pi B_{\rm{ave}}
R^2$, we have
\begin{equation}
{{\cal M}\over{|{\cal W}|}}={5\over{9a\pi^2}}\ {{\Phi^2}\over{GM^2}}\ 
\left(1-{R\over{R_0}}\right)={{2\alpha_{\rm{mag}}}\over{3a\beta}}\ 
\left(1-{R\over{R_0}}\right)\ ,
\label{eq:17}
\end{equation}
so that equation (\ref{eq:11}) gives
\begin{equation}
{1\over{\alpha_{\rm{mag}}}}={1\over{\alpha_{\rm{non}}}}+{{2}\over{3a\beta}}\ 
\left(1-{R\over{R_0}}\right)\ .
\label{eq:18}
\end{equation}
With $B_0\simeq3\ \mu$G and $B_{\rm{ave}}\simeq30-40\ \mu$G for GMCs
(\markcite{my88b}Myers \& Goodman 1988b), flux conservation gives $R/R_0
\simeq0.3$; and ${\cal U}\simeq{\cal M}$ implies $\beta\simeq1$. Thus,
$\alpha_{\rm{mag}}\approx1$ when $\alpha_{\rm{non}}\approx2$, as expected.

Some indication of the reliability of equations (\ref{eq:16}) and
(\ref{eq:17}), and of the approximations leading to them, can be had by
comparing the critical masses they predict for magnetized, isothermal spheres
($a=1.221$) with those obtained from self-consistent, axisymmetric numerical
calculations. In particular, the mass $M_{\Phi}$ which separates magnetically
sub- and supercritical clouds, and for which ${\cal M} =|{\cal W}|$, is
given by equation (\ref{eq:17}) as $0.18\ \Phi/G^{1/2}$, only a 50\%
overestimate of the exact result $M_{\Phi}\simeq0.12\ \Phi/ G^{1/2}$
(\markcite{mou76}Mouschovias \& Spitzer 1976; \markcite{tom88}Tomisaka,
Ikeuchi, \& Nakamura 1988). Further, for $M\ga0.24\ \Phi/G^{1/2}$, the
critical masses we obtain by using equation (\ref{eq:11}) in (\ref{eq:16}a)
lie within a factor 2 of those found by \markcite{tom88}Tomisaka et
al.~(1988; see their eq.~[4.7]). (In fact, our formula is more accurate in
the weak-field limit because it approaches the correct $M_{\rm{crit}}=1.182
\ \sigma^4/(G^3P_s)^{1/2}$ for the nonmagnetic, $\Phi=0$ isothermal sphere.)
And for the equipartition $\beta=1$ seen in GMCs, setting $\alpha_{\rm{mag}}
\approx1$ in equation (\ref{eq:17}) implies $M_{\rm{crit}}\approx(5/6\pi^2)^
{1/2}\Phi/ G^{1/2}$. This is roughly 1.6 times our (approximate) $M_{\Phi}$,
and 2.4 times the exact value, which level of agreement is quite acceptable.
In any case, we are led to expect that critical-mass GMCs and cores are
strongly magnetically supercritical, with $M_{\rm{crit}}\approx2\ M_{\Phi}$
--- a result which has also been argued by \markcite{mck89}McKee (1989) and
\markcite{ber92}Bertoldi \& McKee (1992).

\section{Polytropic Equations of State}

The outwards increase of linewidth both within giant molecular clouds as a
whole (\markcite{lar81}Larson 1981; \markcite{mie94}Miesch \& Bally 1994) and
within individual dense cores (\markcite{ful92}Fuller \& Myers 1992;
\markcite{cas95}Caselli \& Myers 1995), immediately suggests the class of
negative-index polytropes as possible models for these structures.  That is,
if $P\propto\rho^{1+1/N}$, then for a polytropic index $N<-1$ we have
$P/\rho$ increasing for decreasing $\rho$, as required. Such models have
been studied by, e.g., \markcite{via74}Viala \& Horedt (1974) and
\markcite{mal88}Maloney (1988), and we derive the generalized Bonnor-Ebert
relations for them in Appendix C.  Although there are various physical
arguments to support their use (e.g., \markcite{shu72}Shu et al.~1972;
\markcite{dej80}de Jong, Dalgarno, \& Boland 1980; \markcite{mck95}McKee \&
Zweibel 1995), these polytropes turn out to be unlikely descriptions of real
GMCs (and of course, a positive polytropic index is undesirable because it
is inconsistent with a velocity dispersion that increases with radius).

Here we define $n=N/(N+1)$, so that $P\propto\rho^{1/n}$ and $n>1$ for
$N<-1$. From Appendix C, a nonmagnetic, negative-index polytrope which is
truncated anywhere outside of its constant-density central region (inside
which, $a\simeq1$ and $\alpha_{\rm{non}}$ increases without bound towards
$r=0$) will then satisfy
$$\alpha_{\rm{non}}-a={5\over{6}}(4n-3)\ .$$
Thus, an $n>1$ polytrope has $(5/6)(\alpha_{\rm{non}}-a)^{-1}<1$; and since
its density profile is $\rho\propto r^{-p}$, with $p<2$ everywhere for any
$n$ (Appendix C), we also find $\rho_{\rm{ave}}/\rho_s<3$. According to
equation (\ref{eq:14}), then, {\it a truncated, negative-index polytrope
will always be stable, and never critically so} (as was also noted by
\markcite{mal88}Maloney 1988).  Ultimately, the same steady increase of
linewidth with radius which would recommend the polytropic equation of state
in the first place also proves to be its undoing:  the internal pressure
gradient which results is so shallow as to stabilize a cloud under any
external pressure. The concept of a critical mass is then irrelevant for
these models, which casts doubt on their utility in describing real GMCs or
high-mass cores. Any additional support from a mean magnetic field in the
cloud obviously serves only to exacerbate this problem.

Again, this result follows to some extent from the constraint that the
thermal linewidth $\sigma_c$ be fixed during a perturbation of the cloud. By
contrast, both \markcite{via74}Viala \& Horedt (1974) and
\markcite{chi87}Chi\`eze (1987) consider the possibility that the constant
of proportionality in the relation $P\propto\rho^{1/n}$ is invariant, and
find that polytropes can become unstable for certain truncation radii.
Still, our approach is closely related to a point which is {\it independent
of any rule for cloud stability}:

The minimum value of the nonmagnetic virial parameter for a truncated
polytrope is (eq.~[\ref{eq:b7}])
$$\alpha_{\rm{non}}\geq{5\over{2}}\ {{(4n-3)(2n-1)}\over{6n-5}}\ .$$
As usual, this is a lower limit because $\alpha_{\rm{non}}$ can be very
large indeed if the polytrope is truncated at a very small radius. In the
event of equipartition ${\cal U}={\cal M}$ between kinematic and magnetic
energies, equation (\ref{eq:11}) and the identity $\alpha_{\rm{mag}}=2a\ 
{\cal U}/|{\cal W}|$ lead to
$$\alpha_{\rm{mag}}={{\alpha_{\rm{non}}}\over{1+\alpha_{\rm{non}}/(2a)}}=
{{\alpha_{\rm{non}}}\over{1+(1/2)(1-P_s/P_{\rm{ave}})^{-1}}}\ ,$$
in which case equation (\ref{eq:b11}) and $n>1$ finally imply that
$$\alpha_{\rm{mag}}\geq10\ {{(4n-3)(2n-1)}\over{(6n-5)(6n+1)}}>{{10}
\over{7}}\ .$$
Thus, regardless of whether there exist any unstable modes for truncated
polytropes, the virial parameters of such clouds are significantly larger
than the $\alpha_{\rm{mag}}\simeq1$ ($2\ {\cal U}+{\cal W}\approx0$) seen in
GMCs (\markcite{my88b}Myers \& Goodman 1988b) and in the most massive
molecular cores (\markcite{ber92}Bertoldi \& McKee 1992;
\markcite{wil94}Williams et al.~1994, \markcite{wil95}1995), even if
dynamically significant mean magnetic fields are allowed to be present. It
is a property of our specific stability criterion that this fact implies the
absence of critically stable equilibria.

The virial parameters of magnetized polytropes could be reduced to
$\alpha_{\rm{mag}}\simeq1$, for any $n$, if the mean field were such that
${\cal M}\simeq2\ {\cal U}$. In this case, however, using equation
(\ref{eq:b11}) for $P_s/P_{\rm{ave}}$ in the virial theorem (\ref{eq:5})
leads to $0.8<{\cal M}/|{\cal W}|<1$ for $n\geq2$, which is difficult to
reconcile with the rather higher degree of magnetic supercriticality that is
observationally inferred for GMCs and massive cores (${\cal M}/|{\cal W}|
\la0.5$, and $M\simeq2M_{\Phi}$; see \S\S2.2, 2.3).

One consequence of all of this is that the nonthermal linewidths in GMCs,
$\sigma_{\rm{NT}}^2=\sigma_{\rm{ave}}^2-kT/\mu m_H$, cannot be attributed
entirely to the pressure of weakly damped Alfv\'en waves, for which
$P\propto\rho ^{1/2}$ ($n=2$, or $N=-2$; \markcite{mck95}McKee \& Zweibel
1995), and thus $\alpha_{\rm{mag}}\ga1.65$ under magnetic equipartition. It
seems almost certain that Alfv\'en waves do play a significant role in the
support of GMCs and cores (e.g., \markcite{aro75}Arons \& Max 1975;
\markcite{pud90}Pudritz 1990); but, as \markcite{mck95}McKee \& Zweibel
(1995) also note, they cannot be uniquely responsible for their large-scale
stability. In this context, we note that the size-linewidth relation between
clouds can be expressed in terms of a mean magnetic field strength, as in,
e.g., \markcite{my88a}Myers \& Goodman (1988a). Specifically, equation
(\ref{eq:16}d) can be used to write (\ref{eq:16}b) in terms of $P_{\rm{ave}}$
rather than $P_s$, and the definition of $\beta$ (\S2.3) relates
$P_{\rm{ave}}$ to $B_{\rm{ave}}$. Then, with a mean mass per particle $\mu=
2.33$ and a kinetic temperature $T=10$ K, we have
$$\sigma_{\rm{NT}}\simeq0.60\ {\hbox{km\ s}}^{-1}\ 
(\alpha_{\rm{mag}}\beta)^{1/4}
\left({{B_{\rm{ave}}}\over{30\ \mu{\hbox{G}}}}\right)^{1/2}
\left({R\over{1\ {\hbox{pc}}}}\right)^{1/2}.$$
The equality is not quite exact here, because the scalings with
$B_{\rm{ave}}$ and $R$ strictly apply to the total $\sigma_{\rm{ave}}$.
Still, this relation is accurate in the typical case, $\sigma_{\rm{NT}}^2\gg
kT/\mu m_H$; and for $\alpha_{\rm{mag}}=\beta=1$, it is consistent with
available data (see \markcite{my88a}Myers \& Goodman 1988a). Thus, although
\markcite{mou95}Mouschovias \& Psaltis (1995) argue for an interpretation of
this result in terms of Alfv\'en waves, we see here that it is independent
of {\it any} assumptions on the physical origin of the nonthermal motions
in GMCs (in principle, they need not even derive from magnetic fields).

Finally, it is clear that any simple ``mixing'' of two polytropes with
different indices will still preclude the existence of a critical (or
low-$\alpha_{\rm{mag}}$) cloud; even the superposition of an isothermal
part, i.e., $P=C_1\rho+C_2\rho^{1/n}$, only admits one if it is essentially
isothermal anyway ($C_1\gg C_2$). Thus, we now turn to a different equation
of state, in which the gas pressure varies only logarithmically with
density.

\section{The Logotrope}

Many theoretical models of star-forming clouds employ the singular
isothermal sphere, in which $\rho\propto r^{-2}$. However, there is some
evidence that \ion{H}{2} regions are concentrated towards the CO centroids of
their parent GMCs such that their three-dimensional number density is most
consistent with an $r^{-1}$ fall-off (\markcite{wal87}Waller et al.~1987;
\markcite{sco87}Scoville et al.~1987). This suggests the possibility that,
in a heavily smoothed, average sense, the internal density structure of
molecular clouds is essentially $\rho\propto r^{-1}$ (see also
\markcite{sol87}Solomon et al.~1987).  Further, extinction measurements
indicate $\rho\propto r^{-1}$ or so in the outer parts of cores as well
(\markcite{cer85}Cernicharo, Bachiller, \& Duvert 1985;
\markcite{stw90}St\"uwe 1990).

\markcite{liz89}Lizano \& Shu (1989; see also \markcite{geh96}Gehman et
al.~1996) introduced the so-called logotropic equation of state for GMCs, $P=
P_{\rm{iso}}+P_{\rm{turb}}$, with $P_{\rm{iso}}\propto\rho$ and
$P_{\rm{turb}}\propto{\hbox{ln}}(\rho/\rho_{\rm{ref}})$. If the central
linewidth of a cloud is to be entirely thermal in origin, the reference
density in $P_{\rm{turb}}$ must be $\rho_{\rm{ref}}=\rho_c$. However,
$P/\rho$ then {\it decreases} with radius, which is incompatible with the
observations. We suggest that a more complete description of GMCs and cores
is given instead by a ``pure'' logotrope:
\begin{equation}
P=\rho_c\sigma_c^2\left[1+A\ {\hbox{ln}}\left({\rho\over{\rho_c}}\right)
\right]\ ,
\label{eq:41}
\end{equation}
with $A>0$ a parameter to be adjusted. The assumption here is that any
nonthermal motions, which presumably arise from MHD turbulence, add to the
thermal pressure such that both are fully accounted for by the equation of
state (\ref{eq:41}). This relation is shown schematically in Fig.~\ref{fig0},
along with an isothermal-sphere combination and the polytropic
$P\propto\rho^{1/2}$ for Alfv\'en waves. In the rest of our discussion,
equation (\ref{eq:41}) is referred to simply as a logotrope, with the
understanding that no further consideration is given to any mixed-isothermal
version.

\placefigure{fig0}

Given equation (\ref{eq:41}), the equation of hydrostatic equilibrium
(\ref{eq:1}) integrates to
\begin{equation}
{\rho\over{\rho_c}}={1\over{1+\psi/A}}\ ,
\label{eq:42}
\end{equation}
where $\psi=(\phi-\phi_c)/\sigma_c^2$ is a dimensionless gravitational
potential. Substitution of this expression into Poisson's equation
(\ref{eq:2}) shows the existence of a singular solution,
\begin{equation}
{\rho\over{\rho_c}}=\sqrt{{{2A}\over{9}}}\ \left({r\over{r_0}}\right)^{-1}\ ,
\label{eq:43}
\end{equation}
which describes the density profile of the bounded (finite $\rho_c$) solution
at large radii. (The scale radius $r_0=3\sigma_c/[4\pi G\rho_c]^{1/2}$.) In
spite of such a slowly decreasing internal density, a logotrope has only a
finite extent, since for any positive $A$ the pressure will eventually
vanish. Assuming that $A$ is {\it small} enough for the singular density
profile to apply at the edge of the cloud, we have
\begin{equation}
\xi_{\rm{max}}={{R_{\rm{max}}}\over{r_0}}=\sqrt{{{2A}\over{9}}}\ e^{1/A}\ .
\label{eq:44}
\end{equation}
The velocity dispersion, $\sigma^2=P/\rho$, increases with radius inside the
cloud until it reaches a maximum of
\begin{equation}
\sigma_{\rm{max}}^2/\sigma_c^2=Ae^{1/A-1}
\label{eq:44a}
\end{equation}
at $\xi=r/r_0=e^{-1}\xi_{\rm{max}}$, beyond which it falls again (to 0, at
$\xi_{\rm{max}}$).

A logotrope truncated by the pressure $P_s$ of an external medium has
precisely one critical mode for any given value of $A$. As discussed in
Appendix B, if a cloud is truncated at a very small radius $\xi$, it will
always be stable. But in the power-law part of the logotrope, equations
(\ref{eq:42}) and (\ref{eq:43}) show that $d\psi/d\xi=(9A/2)^ {1/2}$, so
once $\xi$ is so large that $P_s/P_c<A/4$, the cloud is unstable
(cf.~eq.~[\ref{eq:a5a}]). Thus,
\begin{equation}
\xi_{\rm{crit}}={{R_{\rm{crit}}}\over{r_0}}=\sqrt{{{2A}\over{9}}}\ 
{\hbox{exp}}\left({1\over{A}}-{1\over{4}}\right)=e^{-1/4}\xi_{\rm{max}}\ ,
\label{eq:45}
\end{equation}
where we have again assumed that $A$ is relatively small and the singular
solution (\ref{eq:43}) holds at $\xi_{\rm{crit}}$. (This expression shows
that $\sigma$ achieves its maximum {\it inside} $\xi_{\rm{crit}}$, and the
surface of the critical logotrope is cooler than its interior.
\markcite{mck89}McKee [1989] has suggested that such behavior might arise
from the radiation of Alfv\'en waves into the ambient medium.) Now,
evaluating the pressure (\ref{eq:41}) and density (\ref{eq:43}) at the
radius (\ref{eq:45}) yields, for any critically stable logotrope,
\begin{equation}
\left({{\rho_s}\over{\rho_c}}\right)_{\rm{crit}}={\hbox{exp}}
\left({1\over{4}}-{1\over{A}}\right)\ \ \ \ \ \ \ \ 
\left({{\sigma_s^2}\over{\sigma_c^2}}\right)_{\rm{crit}}={A\over{4}}\ 
{\hbox{exp}}\left({1\over{A}}-{1\over{4}}\right)\ \ \ \ \ \ \ \ 
\left({{P_s}\over{P_c}}\right)_{\rm{crit}}={A\over{4}}\ .
\label{eq:47}
\end{equation}
Evidently, self-consistency in the use of equation (\ref{eq:43}) here
requires $A\ll4$. Equations (\ref{eq:4d}) and (\ref{eq:6}) can then be used
to find
\begin{equation}
\left({{\rho_{\rm{ave}}}\over{\rho_s}}\right)_{\rm{crit}}={3\over{2}}
\ \ \ \ \ \ \ \ 
\left({{\sigma_{\rm{ave}}^2}\over{\sigma_s^2}}\right)_{\rm{crit}}=
{{14}\over{9}}\ \ \ \ \ \ \ \ 
\left({{P_{\rm{ave}}}\over{P_s}}\right)_{\rm{crit}}={7\over{3}}\ ,
\label{eq:48}
\end{equation}
independently of $A$.

A numerical solution for our logotrope with $A=0.18$ (which we justify
shortly) yields the internal density, pressure, and velocity-dispersion
profiles shown in Fig.~\ref{fig1}. Equations (\ref{eq:44}) and (\ref{eq:45})
are valid here, as an $r^{-1}$ density profile is realized well before the
cloud ends (the vertical line in all four panels of the Figure marks the
radius $\xi_{\rm{crit}}$). It is the steep pressure gradient near
$\xi_{\rm{max}}$ which sets the logotrope apart from the negative-index
polytropes and allows for a critical cloud with a small virial parameter.

\placefigure{fig1}

The {\it total} velocity dispersion $\sigma^2$ in Fig.~\ref{fig1} is roughly
constant near the center of the cloud; and its rise with radius further on is
consistent with $\sigma^2\propto r^{2/3}$, which is essentially the scaling
originally found for GMCs by \markcite{lar81}Larson (1981). Identifying the
central dispersion $\sigma_c$ with the thermal part of the linewidth (so
that thermal motions dominate on the smallest scales, as is observed), the
nonthermal contribution to cloud support is $\sigma_{\rm{NT}}^2=\sigma^2-
\sigma_c^2$, which is also shown in Fig.~\ref{fig1}. At small to moderate
radii, $\sigma_{\rm{NT}}$ naturally rises more steeply than the total
$\sigma$, while at larger $\xi$ the two are more comparable in magnitude.
Roughly, then, our model has $\sigma_{\rm{NT}}^2\propto r$ over some range
in radius before it flattens to $\sigma_{\rm{NT}}^2\propto r^{1/2}$ and
eventually turns over (the maximum $\sigma_{\rm{NT}}^2/\sigma_c^2$ is just
$\sigma_{\rm{max}}^2/\sigma_c^2-1$, and occurs at $\xi=e^{-3/4}
\xi_{\rm{crit}}$). It is found that $\sigma_{\rm{NT}}\propto r^{0.5}$ within
GMCs (e.g., \markcite{\my88b}Myers \& Goodman 1988b), and low-mass molecular
cores show the same general trend; high-mass cores are more suggestive of
$\sigma_{\rm{\rm{NT}}}\propto r^{0.25}$ (\markcite{ful92}Fuller
\& Myers 1992; \markcite{cas95}Caselli \& Myers 1995). A logotropic equation
of state both for individual cores and for entire cloud complexes is
consistent with these observations, provided that low-mass cores can be
viewed as spheres which are pressure-truncated at radii
$\xi<\xi_{\rm{crit}}$. We discuss this point further in \S4.1 below. As
always, though, no matter how $\sigma$ varies with $r$ internally, the {\it
average} relation $\sigma_{\rm{ave}}\propto R^{1/2}P_s^{1/4}$ still holds
for critical-mass clouds, i.e., for GMCs.

A handle on an appropriate value of the parameter $A$ for GMCs as a whole,
can be gotten by reinventing them as smoothed-out spheres with a number of
distinct overdense regions (the cores) sprinkled throughout. The observed
contrasts in mean density and velocity dispersion between GMCs and cores can
then be related to the $A$ of a model logotrope.
\markcite{har94}Harris \& Pudritz (1994; see their Table 2) list each of
$\rho_{\rm{ave}}$, $\sigma_{\rm{ave}}$, and $\Sigma$ for a GMC of mass
$\overline{M}_{\rm{GMC}} =3.3\times10^5\ M_{\odot}$ and a core with
$\overline{M}_{\rm{core}}=5.4\times10^2\ M_{\odot}$. (Such values are
``typical'' in the sense that, {\it by mass}, half of all GMCs [cores] are
larger than $\overline{M}_{\rm{GMC}}$ [$\overline{M}_{\rm{core}}$].)
Defining $\widetilde{\rho}$, $\widetilde {\sigma}$, and $\widetilde{\Sigma}$
as the core-to-GMC ratios of mean volume densities, linewidths, and column
densities, these data show that
\begin{equation}
\widetilde{\rho}\simeq240\ \ \ \ \ \ \ \ 
\widetilde{\sigma}\simeq0.29\ \ \ \ \ \ \ \ 
\widetilde{\Sigma}\simeq4.64\ .
\label{eq:46}
\end{equation}
Cores with $M=\overline{M}_{\rm{core}}$ are among the most massive observed,
and are likely near critical. If they are also logotropes, then their
$\rho_{\rm{ave}}/\rho_s$, etc., should be roughly given by equation
(\ref{eq:48}), regardless of whether $A_{\rm{GMC}}=A_{\rm{core}}$. Such
large cores might further be expected to lie in very dense, cold, highly
pressured regions of their parent clouds. Then the pressure $P_s$ at the
surface of a core is comparable to the GMC's $P_c$, and similarly for the
density and velocity dispersion. These assumptions, together with the
relation $\Sigma\propto P_s^{1/2}$ (eq.~[\ref{eq:16}c]), result in
\begin{equation}
\widetilde{\rho}\simeq\left({{\rho_c}\over{\rho_s}}\right)_{\rm{GMC}}
\ \ \ \ \ \ \ \ 
\widetilde{\sigma}\simeq\left({{\sigma_c}\over{\sigma_s}}\right)_{\rm{GMC}}
\ \ \ \ \ \ \ \ 
\widetilde{\Sigma}\simeq\left({{P_c}\over{P_s}}\right)_{\rm{GMC}}^{1/2}\ .
\label{eq:49}
\end{equation}
Comparison of equations (\ref{eq:46}) and (\ref{eq:47}) then shows that
the observed $\widetilde{\rho}$, $\widetilde{\sigma}$, and $\widetilde
{\Sigma}$ (only two of which are actually independent) are realized in a
logotropic GMC with
\begin{equation}
A_{\rm{GMC}}\simeq0.175\ .
\label{eq:410}
\end{equation}

Although this procedure is highly idealized, it is perhaps telling that an
$A$ exists at all which gives at once the correct $\widetilde {\rho}$ and
$\widetilde {\sigma}$. The logotrope appears to be the simplest barotropic
equation of state which can account for these ratios {\it and} the small
virial parameters (or the critical equilibrium) of GMCs and massive cores.
For example, if GMCs and large cores were both $n=2$ polytropes truncated at
$\xi=r/r_0\simeq21.4$, then equation (\ref{eq:46}) could be roughly
satisfied, but they would also have a large $\alpha_{\rm{non}}=5.37$, and
hence $\alpha_{\rm{mag}}\simeq1.75$ for magnetic equipartition (again,
${\cal M}\simeq2\ {\cal U}$ is required to reduce $\alpha_{\rm{mag}}$ to
unity for a negative-index polytrope, but this is not wholly satisfactory
for other reasons; see \S3). Alternatively, adding an explicit $\rho^{1/2}$
term to equation (\ref{eq:41}) destroys the {\it simultaneous}
correspondence between $\widetilde{\rho}$ and $\widetilde{\sigma}$ for
critical clouds.

To this point, we have not accounted for any mean magnetic fields in GMCs and
cores. However, these should have no effect on our estimate of $A_{\rm{GMC}}$
insofar as very massive clumps have the same $\alpha_{\rm{mag}}\approx1$ as
their parent clouds. More important is the assumption that the cores here
lie at the rare, highest-pressure peaks of cloud complexes; if instead they
are found in less extreme regions, then equation (\ref{eq:410})
is an upper limit to $A_{\rm{GMC}}$. Even so, more direct estimates of
$A_{\rm{GMC}}$ are in fairly good agreement with the representative value
quoted above. For example, equations (\ref{eq:47}) and (\ref{eq:48}) can be
combined to obtain an expression for $\sigma_{\rm{ave}}^2/\sigma_c^2$;
comparison with observed values of the total $\sigma_{\rm{ave}}$ and the
thermal $\sigma_c$ in GMCs then suggests $0.12\la A_{\rm{GMC}}\la0.16$.

On a final note, since $\rho\propto r^{-1}$ at the edges of such logotropes,
we have $a=(1-1/3)/(1-2/5)=10/9$. With the ratio of mean kinematic and
magnetic pressures $\beta\simeq1$, equations (\ref{eq:15}) and (\ref{eq:18})
then give for the critical virial parameters,
$$\alpha_{\rm{non}}=35/18\ \ \ \ \ \ {\hbox{and}}\ \ \ \ \ \ 
\alpha_{\rm{mag}}\simeq1.07.$$
These are near 2 and 1, as expected from general arguments, and not far from
the values 2.054 and 1.15 for a Bonnor-Ebert isothermal sphere.  As alluded
to earlier (\S2.2), the mass-radius and size-linewidth relations among real
GMCs can therefore be understood in terms of this model: equations
(\ref{eq:16}b, c) become
\begin{equation}
{M\over{R^2}}=147\ M_{\odot}\ {\hbox{pc}}^{-2}\ \left({{P_s}\over
{P_{\rm{ISM}}}}\right)^{1/2}\left({{P_{\rm{ISM}}}\over{10^4\ k\ 
{\hbox{cm}}^{-3}\ {\hbox{K}}}}\right)^{1/2}
\label{eq:411}
\eqnum{\ref{eq:411}a}
\end{equation}
and
\begin{equation}
{{\sigma_{\rm{ave}}}\over{R^{1/2}}}=0.37\ {\hbox{km}}\ {\hbox{s}}^{-1}\ 
{\hbox{pc}}^{-1/2}\ \left({{P_s}\over{P_{\rm{ISM}}}}\right)^{1/4}
\left({{P_{\rm{ISM}}}\over{10^4\ k\ {\hbox{cm}}^{-3}\ {\hbox{K}}}}
\right)^{1/4}\ ,
\eqnum{\ref{eq:411}b}
\addtocounter{equation}{+1}
\end{equation} which agree well with observations
(cf.~\markcite{elm89}Elmegreen 1989).
Here $P_s$ is the surface pressure on the {\it molecular part} of the GMC,
while $P_{\rm{ISM}}$ is the total pressure of the hot, intercloud medium.
The ratio of these two is typically around $5-10$, due to the weight placed
on a GMC by its UV-shielding atomic layer (\markcite{elm89}Elmegreen 1989;
also \S2.2 above).

\subsection{Comparison With Data: Molecular Cloud Cores}

As a class, dense molecular cores seem somewhat more heterogeneous than
their parent GMCs. First, many studies (\markcite{car87}Carr 1987;
\markcite{lor89}Loren 1989; \markcite{stt90}Stutzki \& G\"usten 1990;
\markcite{lad91}Lada, Bally, \& Stark 1991) show that the size-linewidth
relation between cores can be significantly weaker than the virial,
$\sigma_{\rm{ave}}\propto R^{1/2}$ scaling among GMCs, and in some cases is
difficult to discern at all. Second, in the L1630 GMC at least, the
size-linewidth relation is considerably better defined for clumps which
stand out more strongly against the interclump medium (see the comparison in
Fig.~13 of Lada et al.~between their ``5$\sigma$'' and ``3$\sigma$'' cores).
And third, the virial parameters $\alpha_{\rm{mag}}$ of GMC cores are not
restricted to values near 1, but rather can range as high as $\approx$100
(\markcite{ber92}Bertoldi \& McKee 1992; \markcite{wil94}Williams et
al.~1994, \markcite{wil95}1995).

This last point is particularly significant because the virial parameters of
cores in a number of different clouds are seen to correlate with their size:
low mass goes along with rather large $\alpha_{\rm{mag}}$, and high mass with
more moderate $\alpha_{\rm{mag}}\approx1$. Thus, as \markcite{ber92}Bertoldi
\& McKee (1992) argue, the smallest clumps in GMCs are far removed from
gravitational instability (surface pressure is more important than
self-gravity in their confinement), while the largest are actually near
criticality (strongly self-gravitating; see also \S2.2 above). Equivalently,
a range in the masses of cores can be viewed, at least roughly, as a range
in the dimensionless radii $\xi$ --- small for low mass, and near
$\xi_{\rm{crit}}$ for high mass --- at which they are truncated by the
ambient pressure in a GMC (virial parameters generally decrease with $\xi$
for any equation of state; Appendix A). This notion could account for the
rather confused size-linewidth relation {\it between} cores (if $\xi$ does
not have a fixed value, then neither do $\alpha_{\rm{non}}$ and
$\alpha_{\rm{mag}}$ in eq.~[\ref{eq:12}b]), and ultimately must also have
consequences for the interpretation of data on the radial variation of
velocity dispersion {\it inside} cores.

If indeed very low-mass cores differ from very high-mass ones mainly in being
truncated at $\xi_{\rm{low}}\ll\xi_{\rm{high}}\simeq\xi_{\rm{crit}}$, then
the former must be less centrally condensed, have ratios $P_s/P_c$ nearer
unity, and be generally less distinguishable from the intercore (GMC) gas
than the latter. This results in the scale radius $r_0=3\sigma_c/(4\pi G
\rho_c)^{1/2}\propto \sigma_c^2/P_c^{1/2}$ being {\it larger} for less
massive clumps. To see this, compare the $r_0$ of a low-mass core, truncated
at $\xi$ small enough that $P_s\approx P_c$, with the $r_0$ of a
near-critical, high-mass core. On average, any two cores should be under
similar surface pressures $P_s$ (this essentially being set by the internal
$P_{\rm{ave}}$ of a parent GMC), so that
\begin{equation}
{{r_0{\rm{(high)}}}\over{r_0{\rm{(low)}}}}\ \simeq\ 
\left({{P_s}\over{P_c}}\right)_{\rm{high}}^{1/2}
\left[{{\sigma_c^2{\rm{(high)}}}\over{\sigma_c^2{\rm{(low)}}}}\right]\ 
\simeq\ \sqrt{{A\over{4}}}\ 
\left[{{\sigma_c^2{\rm{(high)}}}\over{\sigma_c^2{\rm{(low)}}}}\right]\ ,
\label{eq:412}
\end{equation}
where the second step follows from equation (\ref{eq:47}) if a logotropic
equation of state applies. Thus, for $\sigma_c$ not too widely different
between the two cores, and for $A$ small enough, the ratio (\ref{eq:412})
will be $<1$. The same {\it physical} radius $r$ in high- and low-mass cores
then corresponds to respectively larger and smaller $\xi=r/r_0$. This result,
which just reflects the fact that more strongly self-gravitating structures
are more centrally concentrated, must be considered when attempting to match
any model to any observations.

Returning now to the issue of internal velocity-dispersion profiles,
\markcite{ful92}Fuller \& Myers (1992; hereafter FM) and
\markcite{cas95}Caselli \& Myers (1995; hereafter CM) have compiled linewidth
measurements in at least three different molecular lines for each of 14
low-mass and 24 massive cores. The resulting $\sigma$ vs.~$r$ profiles
indicate that, over similar ranges $r\sim0.1-1$ pc in all the clumps, and
independently of whether or not stars are present within them, the velocity
dispersion rises more steeply with radius in low-mass cores than in
high-mass ones. In particular, nonthermal linewidths in the former are
consistent with $\sigma_{\rm{NT}}\propto r^{0.53}$; in the latter,
$\sigma_{\rm{NT}}\propto r^{0.21}$ (\markcite{cas95}CM). In our view, this
situation is a direct result of the effect summarized by equation
(\ref{eq:412}). That is, we consider the low-mass cores of \markcite{ful92}FM
to be eminently stable logotropes truncated at small $\xi$, while the
high-mass clumps of \markcite{cas95}CM are near critical stability and hence
have smaller $r_0$. The two surveys target similar $r$ within each type of
core, so the massive-core data apply to larger dimensionless $\xi$, and must
sample the shallower parts of velocity-dispersion profiles like those in
Fig.~\ref{fig1}.

To actually confront the predictions of our logotropic equation of state with
the observations of \markcite{ful92}FM and \markcite{cas95}CM, we again
identify the (model) central velocity dispersion with the (observed) thermal
linewidth: $\sigma_c^2=\sigma_{\rm{T}}^2=kT/\mu m_H$, where $T$ is taken
from \markcite{cas95}CM, and $\mu=2.33$. \markcite{cas95}CM assume $T=10$ K
for all the low-mass cores, and we further assume them all to have
$P_s\approx P_c$; they should therefore have a common $r_0$, which may be
easily estimated. Five of the \markcite{ful92}FM cores are observed at radii
$r_{\rm{TNT}}$ such that $\sigma_{\rm{NT}}=\sigma_{\rm{T}}$; among these,
$\langle r_{\rm{TNT}}\rangle\simeq0.12$ pc (see Table 3 of
\markcite{cas95}CM). The related quantity $\xi_{\rm{TNT}}=r_{\rm{TNT}}/r_0$
in a model logotrope is just that point where $\sigma^2/\sigma_c^2=
(\sigma_{\rm{NT}}^2+\sigma_{\rm{T}}^2)/\sigma_{\rm{T}}^2=2$. If, for example,
$A=0.2$ in equation (\ref{eq:41}), we find $\xi_{\rm{TNT}}=0.51$, and thus
$$r_0{\hbox{(low)}}\approx0.25\ {\hbox{pc}}\ .$$
This result is then used in equation (\ref{eq:412}), along with $\sigma_c^2
\propto T$ and $A=0.2$, to give
$$r_0{\hbox{(high)}}\approx0.056\ {\hbox{pc}}\ \left({{T_{\rm{high}}}\over
{10\ {\hbox{K}}}}\right)\ ,$$
where $12\ {\hbox{K}}\leq T\leq33\ {\hbox{K}}$ for the massive cores
(\markcite{cas95}CM). The data may now be appropriately scaled and compared
to theoretical $\sigma-r$ and $\sigma_{\rm{NT}}-r$ curves.

Figure \ref{fig2} is the result of such a comparison with three logotropes
of different $A$. The good overall agreement between the models and data
suggests that GMC cores are well described by
$$A_{\rm{core}}=0.20\pm0.02\ ,$$
which is encouragingly close to $A_{\rm{GMC}}$ as given in equation
(\ref{eq:410}) above. In Table \ref{tbl1} we list the radii and other
properties (as calculated from eqs.~[\ref{eq:44}] to [\ref{eq:47}] above) of
critical logotropes with $A$ near 0.2.

\placefigure{fig2}

\placetable{tbl1}

Some of the scatter in Fig.~\ref{fig2} could arise if $A$ does not have a
common value for all cores, and any embedded stars could further complicate
the situation. (Notably, \markcite{ful92}FM and \markcite{cas95}CM both find
that starless cores alone show a tighter --- but not different ---
linewidth-radius relation than do cores with stars, suggesting that the
basic form seen here is one of the initial conditions for star formation.)
However, some of the scatter is surely due to the fact that the cores in
this sample do not reside in a single GMC, but come from diverse
environments. (For instance, the low-mass cores here tend to be found in dark
clouds of mass $\sim10^4-10^5M_\odot$, and the high-mass ones in somehat
larger GMCs.) Thus, there is no guarantee that every low-mass core observes
$P_s\approx P_c$, or that every high-mass core is at its critical mass. It
is the large number of cores available here which allows the mean trend in
Fig.~\ref{fig2}, and the implied structural dichotomy between small and large
clumps, to emerge.

As with our earlier consideration of core-to-GMC density and linewidth
ratios, a fit of similar quality to the data in Fig.~\ref{fig2} can be
obtained with the model velocity-dispersion profile of an $n=2$ polytrope
which has a scale $r_0$ about a factor of two smaller than that used here.
As we have stressed, however, there are other difficulties with the
application of polytropic equations of state to GMCs and cores.

In summary, the approach we have taken self-consistently allows for purely
thermal motions to dominate on the smallest scales in cores (and in GMCs);
for an internal density structure ($\rho\propto r^{-1}$ in the outer parts of
cores, and much shallower near their centers) which is consistent with at
least some observations\footnotemark\footnotetext{\markcite{wil95}Williams et
al.~(1995) fit the projected density profiles of many clumps in the Rosette
GMC, with a function of the form $(1+r_p/a)^{-n}$ --- $a$ being the projected
half-power radius. They find $n\approx1$ outside of unresolved central
regions, and suggest that this implies an intrinsic density profile of
$\rho\propto r^{-2}$. However, because molecular cores have a finite extent,
their density profiles in projection are not so simply related to their
space densities. (For example, the surface density of a core with {\it any}
$\rho$ profile must decline steeply in the outermost regions, as less and
less of the core material is intercepted by the line of sight.) Even a
truncated logotrope, for which we might naively expect a flat surface-density
profile, actually gives rise to something consistent with $n\approx-0.8$ to
$-1.0$ over most radii in the fitting function of \markcite{wil95}Williams et
al.; and the projected half-power radius is only $a=2.92r_0\sim0.2$ pc
for a critical, $A=0.2$ core ($R_{\rm{crit}}=24.37r_0$).}
(\markcite{cer85}Cernicharo et al.~1985; \markcite{stw90}St\"uwe 1990;
\markcite{and93}Andr\'e, Ward-Thompson, \& Barsony 1993); and for a unified
treatment of low- and high-mass cores. This is in contrast to the models of
\markcite{ful92}FM and \markcite{cas95}CM (also \markcite{mye92}Myers \&
Fuller 1992), which assume two distinct components of isothermal and
nonthermal gas in cores and predict a $\rho\propto r^{-2}$ singularity at
their centers. It should also be noted that the assumption $P_s\approx P_c$
in low-mass cores still allows for these noncritical clumps to show a fair
degree of central concentration. For example, an $A=0.2$ logotrope which is
truncated at $\xi\simeq0.1\ \xi_{\rm{crit}}$ has $\rho_s/\rho_c<0.09$, but
$P_s/P_c\simeq0.5$; thanks to the relatively weak dependence of $P$ on
$\rho$, we do not require, nor even expect, that all low-mass cores be
uniform-density spheres. Our model is therefore consistent with the
significantly non-uniform density profiles of low-mass cores
($\alpha_{\rm{mag}}\ga3$) in the Rosette GMC (\markcite{wil95}Williams et
al.~1995).

\section{Summary}

We have generalized the classic Bonnor-Ebert relations for isothermal
spheres, to give expressions for the masses, radii, and total velocity
dispersions of magnetized, pressure-truncated gas clouds with any internal
pressure-density relation. The analysis combines an exact approach, based on
solving the equation of hydrostatic equilibrium, with a virial-theorem
treatment to incorporate mean magnetic fields into the clouds. A stability
criterion has been developed that relies only on the assumption of an
invariant central velocity dispersion $\sigma_c$, and is independent of the
gas equation of state. Clouds which are just {\it critically} stable under
this condition have mass-radius and size-linewidth relations which are
effectively oblivious to their internal structure, i.e., to their equation
of state.

These results have been applied to molecular cloud complexes and dense cores,
leading to three main conclusions:

(1) In spite of their nonspherical geometry, the gross features of GMCs are
consistent with those of critical-mass clouds in our analysis. (In the
context of GMCs and cores, we identify the central velocity dispersion with
the {\it thermal} part of the total [thermal plus turbulent] linewidth.)
{\it Independently of the exact equation of state}, if equipartition between
magnetic and kinetic energies obtains in a critically stable cloud, then it
will appear to be in simple gravitational equilibrium, $2\ {\cal U}+{\cal W}
\approx0$ (equivalently, it will have a virial parameter $\alpha_{\rm{mag}}
\simeq1$). These are observed features of real GMCs and their most massive
cores. In addition, any of our critical-mass spheres obey the same
mass-radius and size-linewidth relations which hold among real GMCs. All of
this implies that GMCs, and large clumps, are near criticality.

(2) No polytropic (power-law) pressure-density relation can fully
characterize GMCs or cores. A positive polytropic index leads to a velocity
dispersion which decreases outwards within a cloud, opposite to what is seen
both in individual cores and in entire cloud complexes. Alternatively, a
negative index (i.e., $P\propto\rho^{\gamma}$ with $\gamma<1$) gives a
linewidth which increases with radius, as required. However, such clouds do
not show virial parameters near unity unless their mean fields are so strong
that their magnetic and gravitational energies are comparable in magnitude,
and this is again in conflict with observations of GMCs and large cores.
Given our stability criterion, an equivalent statement is that
pressure-truncated polytropes are unconditionally stable against
gravitational collapse. Since weakly damped Alfv\'en waves satisfy
$P\propto\rho^{1/2}$ (\markcite{mck95}McKee \& Zweibel 1995), this shows
explicitly that they cannot fully explain the observed nonthermal support in
molecular clouds.

(3) The most successful model for the internal structure of molecular cores,
and one which is also consistent with global properties of GMCs, is a ``pure
logotrope:'' $P/P_c=1+A\ {\hbox{ln}}(\rho/\rho_c)$. This relation is meant to
account for {\it all} contributions to the total gas pressure, including the
effects of disordered magnetic fields (MHD turbulence). Once it is recognized
that low-mass cores are far below their critical masses (though still in
virial equilibrium; they are essentially pressure-confined), and that
high-mass cores are near criticality, the internal velocity-dispersion
profiles of clumps from a variety of environments are seen to be consistent
with a logotropic model with $A\simeq0.2$. A similar value of $A$ can also
explain characteristic core-to-GMC ratios of mean densities and linewidths,
and any $A$ is consistent with the observed size-linewidth relation between
GMCs because the logotropic equation of state allows for critical
equilibria. Finally, the equilibrium density profile of a logotrope has
$\rho\propto r^{-1}$, outside of a constant-density central region. There
is some observational support to be found for this prediction, in GMCs and
cores both.

\acknowledgments

We would like to thank Phil Myers for his comments on an earlier version of
this paper, and Charles Curry for valuable discussions. R.E.P.~also
acknowledges conversations with Paola Caselli and Gary Fuller. This work was
supported in part by the Natural Sciences and Engineering Research Council
of Canada.

\appendix

\section{PRESSURE-TRUNUCATED EQUILIBRIA}

For a spherical cloud in hydrostatic equilibrium, with kinematic (thermal
plus turbulent) pressure $P=\rho\sigma^2$ (so that $\sigma$ is the
one-dimensional velocity dispersion) and a self-gravitational potential
$\phi$, we define
$$\psi={{\phi-\phi_c}\over{\sigma_c^2}}, \ \ \ \ \ r_0^2={{9\sigma_c^2}\over
{4\pi G\rho_c}}, \ \ \ \ \ {\hbox{and}} \ \ \ \ \ \xi=r/r_0,$$
where a subscript $c$ denotes evaluation of a quantity at the cloud center.
The characteristic scale of a cloud is set by $r_0$; the factor of 9 in
its definition identifies it with the projected half-power radius of an
isothermal sphere (e.g., \markcite{bin87}Binney \& Tremaine 1987). The
equation of hydrostatic equilibrium then reads
\begin{equation}
{d\over{d\xi}}\left({P\over{P_c}}\right)=-{\rho\over{\rho_c}}\dpsi\ ,
\label{eq:1}
\end{equation}
and Poisson's equation becomes
\begin{equation}
{1\over{\xi^2}}{d\over{d\xi}}\left(\xi^2\dpsi\right)=9{\rho\over{\rho_c}}\ .
\label{eq:2}
\end{equation}
(Note that $\psi=d\psi/d\xi=0$ at $\xi=0$, and $\psi>0$ for $\xi>0$.) The
mass enclosed within radius $\xi$ is given by
$$M\equiv M(\xi)=4\pi \rho_cr_0^3\int_0^{\xi}{\xi^\prime}^{2}{\rho\over
{\rho_c}}\ d\xi^\prime\ ,$$
so that, with the help of equation (\ref{eq:2}),
\begin{equation}
M={{4\pi}\over{9}}\rho_c r_0^3\left(\xi^2\dpsi\right)={{\sigma_c^2 r_0}\over
{G}}\left(\xi^2\dpsi\right)\ .
\label{eq:3}
\end{equation}

Following \markcite{ebe55}Ebert (1955) and \markcite{bon56}Bonnor (1956), the
surface of the cloud is defined by that radius $\xi$ at which the internal
pressure $P(\xi)$ just equals a confining pressure $P_s$ due to a surrounding
medium of negligible gravity. Equation (\ref{eq:3}), the definition of $r_0$,
and the identity $P_c=\rho_c\sigma_c^2$ then yield, for any equation of state
relating $P$ and $\rho$,
\begin{equation}
M=\sqrt{9\over{4\pi}}\left(\xi^2\dpsi\right)\left({P_s\over{P_c}}\right)
^{1/2}{{\sigma_c^4}\over{(G^3P_s)^{1/2}}}\ .
\label{eq:4a}
\end{equation}
The physical radius of the cloud is $R=\xi r_0$:
\begin{equation}
R=\sqrt{{9\over4\pi}}\ \xi\left({P_s\over{P_c}}\right)^{1/2}{{\sigma_c^2}
\over{(GP_s)^{1/2}}}\ .
\label{eq:4b}
\end{equation}
These relations combine to give
\begin{equation}
\Sigma\equiv{M\over{\pi R^2}}=\sqrt{4\over{9\pi}}\ \dpsi\left({P_s\over
{P_c}}\right)^{-1/2}\left({P_s\over{G}}\right)^{1/2}
\label{eq:4c}
\end{equation}
and
\begin{equation}
\rho_{\rm{ave}}\equiv{{3M}\over{4\pi R^3}}={1\over{3}}\left({1\over{\xi}}
\dpsi\right)\left({P_s\over{P_c}}\right)^{-1}{P_s\over{\sigma_c^2}}\ .
\label{eq:4d}
\end{equation}
Given an equation of state, the integration of equations (\ref{eq:1}) and
(\ref{eq:2}) determines $d\psi/d\xi$ and $P_s/P_c$ at any truncation radius
$\xi$. This fixes the coefficients in equations (\ref{eq:4a})--(\ref{eq:4d})
and affords mass-radius and size-linewidth relations for pressure-truncated
clouds.  (Note that the scalings in these relations ---
$M\propto\sigma^4/P_s^{1/2}$, $R\propto\sigma^2/P_s^{1/2}$, etc. --- are
{\it global} ones, and do not necessarily reflect the radial dependence of
any quantity {\it inside} a cloud.)

The definition of the non-uniformity parameter $a$ (eq.~[\ref{eq:9}]) allows
it to be found as a function of radius:
\begin{equation}
a={{15}\over{\xi^3}}\left({{d\psi}\over{d\xi}}\right)^{-2}\int_0^{\xi}
{\xi^\prime}^{3}{{d\psi}\over{d\xi^\prime}}{\rho\over{\rho_c}}\ 
d\xi^\prime\ ;
\label{eq:13a}
\end{equation}
and equation (\ref{eq:3}) implies that the virial parameter $5\sigma_
{\rm{ave}}^2R/GM$ for nonmagnetic clouds (i.e., $\alpha_{\rm{mag}}=
\alpha_{\rm{non}}$) is just
\begin{equation}
\alpha_{\rm{non}}=5{\sigma_{\rm{ave}}^2\over{\sigma_c^2}}
\left(\xi\dpsi\right)^{-1}\ ,
\label{eq:13b}
\end{equation}
which tends to be a nonincreasing function of $\xi$. Alternatively, equations
(\ref{eq:4d}) and (\ref{eq:12}d) give
\begin{equation}
\alpha_{\rm{non}}-a=15\left(\dpsi\right)^{-2}\left({P_s\over{P_c}}\right)\ .
\label{eq:13}
\end{equation}
These results can be used to rewrite equations (\ref{eq:4a})--(\ref{eq:4d})
in terms of $\alpha_{\rm{non}}$, $a$, and $\sigma_{\rm{ave}}$; equations
(\ref{eq:12}) (with $\alpha_{\rm{mag}}=\alpha_{\rm{non}}$) are then
obtained. Also, once $\alpha_{\rm{non}}$ is known as a function of radius in
any cloud, the observable $\alpha_{\rm{mag}}$ for a given magnetic field
configuration may be calculated (e.g., \S2.3).

\section{STABILITY CRITERION}

We begin by recalling the definition of the scale radius $r_0$ from
equation (\ref{eq:3}):
\begin{equation}
r_0={{GM}\over{\sigma_c^2}}\left(\xi^2\dpsi\right)^{-1};
\label{eq:a1}
\end{equation}
and rearranging equation (\ref{eq:4a}):
\begin{equation}
P_s={9\over{4\pi}}\left(\xi^2\dpsi\right)^2\left({P_s\over{P_c}}\right)
{{\sigma_c^8}\over{G^3M^2}}\ .
\label{eq:a2}
\end{equation}
The stability of a cloud truncated at radius $R$ by the external pressure
$P_s$ is determined by the sign of the derivative $(\partial P_s/\partial
R)$: this is negative for a stable equilibrium, 0 for the ``critical''
cloud, and positive for instability.

We now apply a perturbation $R\rightarrow R+dR$, while keeping the mass $M$
and central (i.e., thermal) velocity dispersion $\sigma_c$ of the cloud
fixed, so that
\begin{equation}
\dpdr=\dpdxi\dxidr.
\label{eq:a3}
\end{equation}
Then $R=\xi r_0$ and equation (\ref{eq:a1}) give us
\begin{eqnarray}
\dxidr & = & {1\over{r_0}}\left[1+{\xi\over{r_0}}\left({{\partial r_0}\over
{\partial\xi}}\right)_{M,\sigma_c}\right]^{-1} \nonumber \\
 & = & {1\over{r_0}}\left[1-\xi\left(\xi^2\dpsi\right)^{-1}
{d\over{d\xi}}\left(\xi^2\dpsi\right)\right]^{-1} \nonumber \\
 & = & {1\over{r_0}}\left[1-9\xi\left(\dpsi\right)^{-1}{{\rho_s}\over{\rho_c}}
\right]^{-1}, \label{eq:a4}
\end{eqnarray}
where $\rho_s$ is the density just inside the cloud edge and we have used
Poisson's equation (\ref{eq:2}). We also have, from equation (\ref{eq:a2}),
\begin{eqnarray}
\dpdxi & = & P_s\left(\dpsi\right)^{-1}\left({{P_c}\over{P_s}}\right)\left[
\left(2\d2psi+{4\over{\xi}}\dpsi\right)\left({{P_s}\over{P_c}}\right) +
\left(\dpsi\right)^2{d\over{d\psi}}\left({{P_s}\over{P_c}}\right)\right]
\nonumber \\
 & = & P_s\left(\dpsi\right)^{-1}\left({{\rho_s}\over{\rho_c}}\right)\left[
18-\left(\dpsi\right)^2\left({{P_s}\over{P_c}}\right)^{-1}\right],
\label{eq:a5}
\end{eqnarray}
where both equations (\ref{eq:1}) and (\ref{eq:2}) have been used.

Thus, equations (\ref{eq:a3}), (\ref{eq:a4}), and (\ref{eq:a5}) show that
\begin{equation}
\dpdr=18{{P_s}\over{R}}\left[{{1-(1/18)(d\psi/d\xi)^2(P_s/P_c)^{-1}}\over
{(1/\xi)(d\psi/d\xi)(\rho_c/\rho_s)-9}}\right].
\label{eq:a5a}
\end{equation}
Now, from equations (\ref{eq:13}) and (\ref{eq:4d}), we have
$${1\over{18}}\left(\dpsi\right)^2\left({{P_s}\over{P_c}}\right)^{-1}=
{5\over{6}}{1\over{\alpha_{\rm{non}}-a}}\ \ \ \ \ {\hbox{and}}\ \ \ \ \ 
{1\over{\xi}}\left(\dpsi\right)=3{{\rho_{\rm{ave}}}\over{\rho_c}}\ ,$$
so that finally,
\begin{equation}
\dpdr=-6{{P_s}\over{R}}\left[{{1-(5/6)(\alpha_{\rm{non}}-a)^{-1}}\over
{3-\rho_{\rm{ave}}/\rho_s}}\right].
\label{eq:a7}
\end{equation}
Alternatively, for a {\it nonmagnetic} cloud, we can make use of
equation (\ref{eq:12}c), with $\alpha_{\rm{mag}}=\alpha_{\rm{non}}$, to write
\begin{equation}
\dpdr=-6{{P_s}\over{R}}\left[{{1-(\pi G\Sigma^2)/(8P_s)}\over
{3-\rho_{\rm{ave}}/\rho_s}}\right].
\label{eq:a6}
\end{equation}
This treatment is valid for {\it any} gas equation of state, so long as the
cloud radius is perturbed in such a way that $M$ and $\sigma_c$ are
unaffected. The result (\ref{eq:a6}) was also obtained by
\markcite{ebe55}Ebert (1955) and \markcite{bon56}Bonnor (1956), who
specifically considered the stability of truncated isothermal spheres, and by
\markcite{mal88}Maloney (1988) in his study of negative-index polytropes.

Any virialized cloud truncated at small $\xi$ (i.e., where $\rho$ and
$\sigma^2$ are approximately constant) will be stable. This is because,
regardless of the equation of state, the potential $\psi$ at small radii can
always be expanded in a power series of the form: $\psi=(3/2)\xi^2 -{\cal
O}(\xi^4)$ (see Appendix C for an example). Thus, referring to equation
(\ref{eq:a5a}) with $\rho_s\approx\rho_c$, we have $(\partial P_s/\partial
R)_{M,\sigma_c}\approx-3P_s/R<0$. Depending on the equation of state, there
may or may not be a truncation at larger $\xi$ which results in $(\partial
P_s/\partial R)_{M,\sigma_c}>0$ and an unstable equilibrium.

\section{POLYTROPES OF NEGATIVE INDEX}

Consider the (nonmagnetic) equation of state
$$P\propto\rho^{1+1/N},$$
where the polytropic index $N<-1$. Then we define $n=N/(N+1)$ and write
\begin{equation}
P=\rho_c\sigma_c^2\left({\rho\over{\rho_c}}\right)^{1/n},\ \ \ \ \ \ \ \ 
n\geq1,
\label{eq:b1}
\end{equation}
with $n=1$ corresponding to isothermality.  In order to evaluate the
coefficients in equations (\ref{eq:4a})--(\ref{eq:4d}) (or [\ref{eq:12}],
with $\alpha_{\rm{mag}}=\alpha_{\rm{non}}$), we first use the relation
(\ref{eq:b1}) to integrate the equation of hydrostatic equilibrium
(\ref{eq:1}) and find
\begin{equation}
{\rho\over{\rho_c}}=[1+(n-1)\psi]^{n/(1-n)},
\label{eq:b2}
\end{equation}
so that Poisson's equation (\ref{eq:2}) becomes
\begin{equation}
{1\over{\xi^2}}{d\over{d\xi}}\left(\xi^2\dpsi\right)=
9[1+(n-1)\psi]^{n/(1-n)}.
\label{eq:b3}
\end{equation}
(Note that in the limit $n\rightarrow1$, the right-hand side of
eq.~[\ref{eq:b3}] becomes $9e^{-\psi}$, just what is required for
the isothermal sphere.) At the cloud center, $\psi=d\psi/d\xi=0$, so for
small $\xi$ the potential $\psi$ can be expanded in a power series,
\begin{equation}
\psi\longrightarrow{3\over{2}}\xi^2-{{3n}\over{40}}\xi^4+{\cal O}(\xi^6),
\label{eq:b4}
\end{equation}
allowing for an integration of equation (\ref{eq:b3}). The result of this
is a core-halo structure to the cloud: at very small radii, the solution is
indistinguishable from that for an isothermal sphere, and the density and
velocity dispersion are essentially constant with $\xi$, independently of
$n$. Once $\xi\ga$ a few, however, a power-law density profile obtains; for
each $n$, there exists a singular solution to which the equilibrium cloud
tends at large radii.

These singular solutions are readily found by substituting the prescription
$${\rho\over{\rho_c}}=K\xi^{-p}$$
into equation (\ref{eq:b2}) and solving (\ref{eq:b3}) for $K$ and $p$. This
gives
\begin{equation}
p={{2n}\over{2n-1}}\ \ \ \ \ \ \ {\hbox{and}}\ \ \ \ \ \ \ 
K=\left[{2\over{9}}\ (p-1)(3-p)\right]^{p/2},
\label{eq:b5}
\end{equation}
so that with $n\geq1$ we have $1<p\leq2$, and these clouds are diffuse
enough that they have no ``natural'' edge. That is, the density never
vanishes, and any boundary to the cloud must be defined by the point at
which the internal pressure matches that of an external medium. (It is again
worth noting the exact correspondence to the singular isothermal sphere when
$n=1$ in eq.~[\ref{eq:b5}]; cf.~\markcite{cha67}Chandrasekhar 1967.)

Consider now a polytrope of negative index, which is truncated at radius
$\xi$ large enough that the structure of the cloud is given by the singular
solution. Then $d\psi/d\xi$ and $P_s/P_c$ can be calculated analytically, and
equation (\ref{eq:13}) yields
\begin{equation}
\alpha_{\rm{non}}-a={5\over{6}}(4n-3)\ .
\label{eq:b6}
\end{equation}
We also know that $a=(1-p/3)/(1-2p/5)$, so
\begin{equation}
\alpha_{\rm{non}}={5\over{2}}\ {{(4n-3)(2n-1)}\over{6n-5}}
\label{eq:b7}
\end{equation}
and we have the following:
\begin{equation}
M=\sqrt{{2\over{\pi(4n-3)^3}}}\ \left({{6n-5}\over{2n-1}}\right)^2\ 
{{\sigma_{\rm{ave}}^4}\over{(G^3P_s)^{1/2}}}\ ,
\label{eq:b8}
\end{equation}
\begin{equation}
R=\sqrt{1\over{2\pi(4n-3)}}\ {{6n-5}\over{2n-1}}\ 
{{\sigma_{\rm{ave}}^2}\over{(GP_s)^{1/2}}}\ ,
\label{eq:b9}
\end{equation}
\begin{equation}
\Sigma=\sqrt{8\over{\pi(4n-3)}}\ \left({{P_s}\over{G}}\right)^{1/2}\ ,
\label{eq:b10}
\end{equation}
and
\begin{equation}
\rho_{\rm{ave}}={{6n-3}\over{6n-5}}\ {{P_s}\over{\sigma_{\rm{ave}}^2}}\ .
\label{eq:b11}
\end{equation}
Thus, regardless of where they are truncated by the surface pressure $P_s$,
these model clouds can only ever represent modest enhancements over the
intercloud medium. In particular, $P_{\rm{ave}}/P_s$ has a maximum of 3 (for
isothermality, $n=1$), but even for $n=2$ (which gives the $\rho$ dependence
of Alfv\'en wave pressure; \markcite{mck95}McKee \& Zweibel 1995) is reduced
to just 9/7.

As discussed in \S3, any truncation of a negative-index polytrope results in
a stable cloud. This is because, with $\rho\propto r^{-p}$, we have that
$\rho_{\rm{ave}}/\rho_s=3/(3-p)$, and the stability criterion (\ref{eq:a7})
becomes (with the help of [\ref{eq:b6}]) $(\partial P_s/\partial
R)_{M,\sigma_c}=-4P_s/R<0$. In particular, this holds in the limit
$n\rightarrow1$, when the numerator and denominator of (\ref{eq:a7}) both
vanish. What allows for the existence of critical and unstable equilibria for
the bounded isothermal sphere, then, is the fact that its density profile
actually oscillates about the singular solution $\rho\propto r^{-2}$ at large
radii (\markcite{cha67}Chandrasekhar 1967). This does {\it not} occur for
those polytropes with $n>1$; instead, the bounded spheres follow the singular
solutions exactly at large $\xi$.

Finally, it is worth noting that these results do not contradict the well
known instability (e.g., \markcite{shu77}Shu 1977) of truly singular
spheres, for which $\rho\propto r^{-p}$ all the way to the center and
$\rho_c\rightarrow\infty$. In such cases, perturbation of the truncated
cloud radius is performed with $M$, $\sigma_c$, {\it and} $\rho_c$ all held
fixed. Stability then depends only on the sign of $(\partial P_s/\partial\xi)
_{M,\sigma_c}$ (eq.~[\ref{eq:a5}]), which is always positive for these
polytropes.

\clearpage

\begin{deluxetable}{ccccccc}
\tablecaption{Properties of Critical Logotropes. \label{tbl1}}
\tablewidth{0pt}
\tablehead{
\colhead{$A$} & \colhead{$\xi_{\rm{max}}$} & \colhead{$\xi_{\rm{crit}}$} &
\colhead{$\rho_s/\rho_c$} & \colhead{$\sigma_s^2/\sigma_c^2$} &
\colhead{$\sigma_{{\rm{NT}},s}^2/\sigma_c^2$} & \colhead{$P_s/P_c$}
}
\startdata
0.18 & 51.73 & 40.29 & $4.96\times10^{-3}$ & 9.07 & 8.07 & 0.045 \nl
0.20 & 31.29 & 24.37 & $8.65\times10^{-3}$ & 5.78 & 4.78 & 0.050 \nl
0.22 & 20.83 & 16.22 & $1.36\times10^{-2}$ & 4.04 & 3.04 & 0.055 \nl
\enddata
\end{deluxetable}

\clearpage

\figcaption[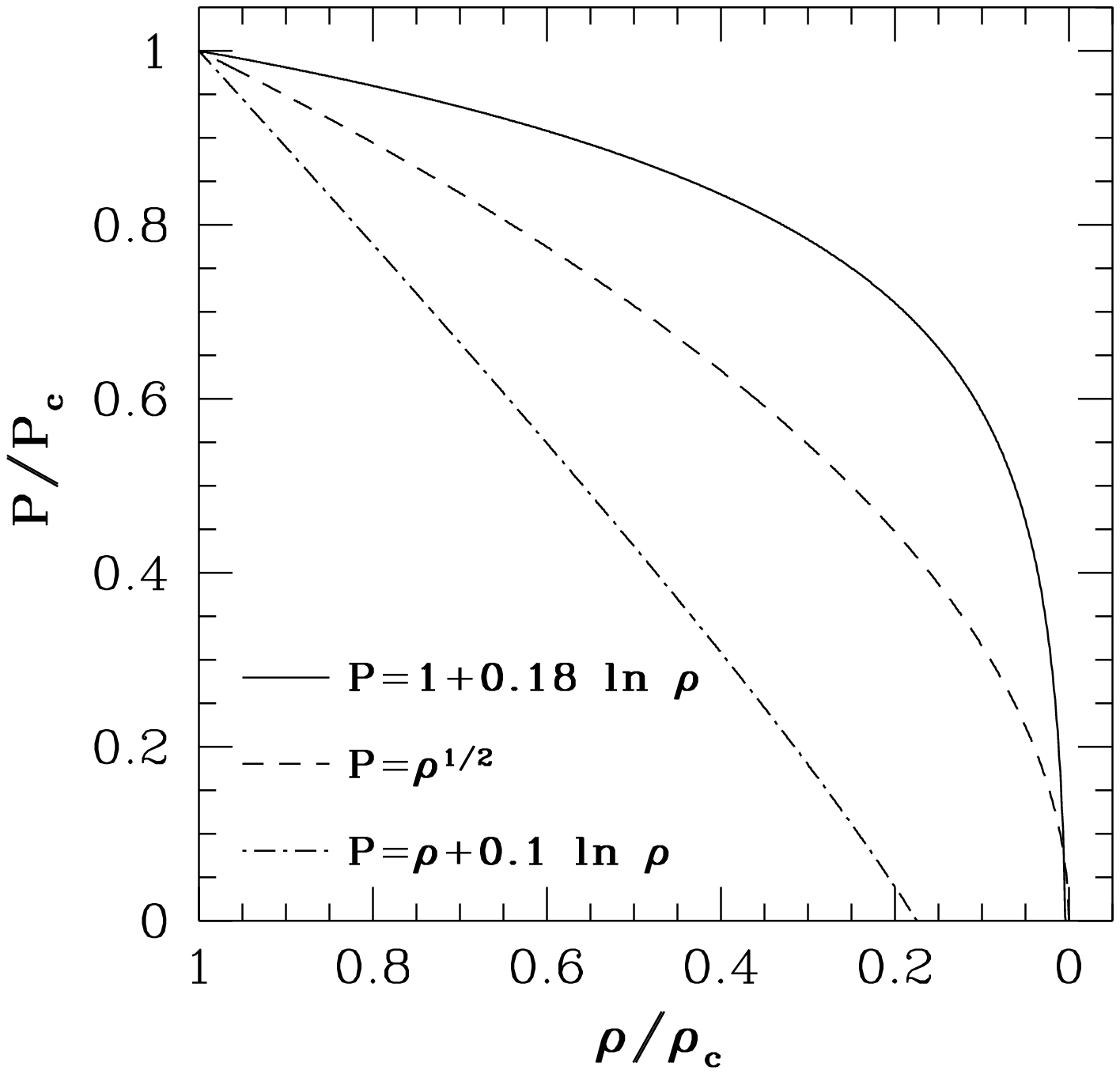]{Comparison of the logotropic equation of state
[$P/P_c=1+0.18\ {\hbox{ln}}(\rho/\rho_c)$] with those for Alfv\'en wave
pressure [$P/P_c=(\rho/\rho_c)^{1/2}$] and a logotrope-plus-isothermal sphere
combination [$P/P_c=\rho/\rho_c+0.1\ {\hbox{ln}}(\rho/\rho_c)$].
\label{fig0}}

\figcaption[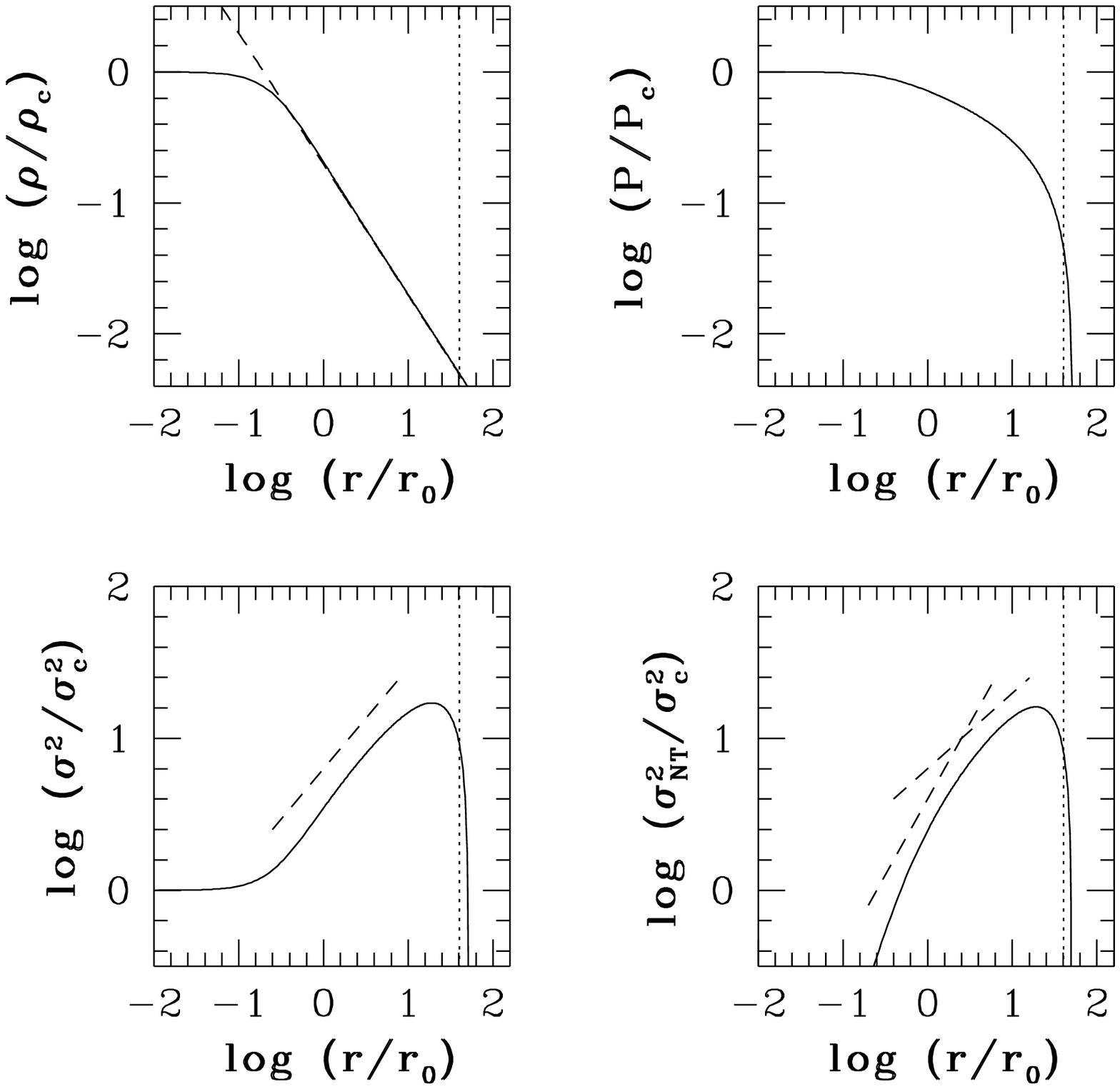]{Structure of a logotropic gas sphere with $A=0.18$.
The vertical line in all four panels is at the truncation radius which makes
for a critically stable cloud: $\xi_{\rm{crit}}=40.29$. The long-dashed line
in the plot of $\rho$ vs.~$r$ is the singular density profile of eq.~(4.3);
that in the plot of total linewidth (bottom left) represents $\sigma^2\propto
r^{2/3}$; and those in the plot of nonthermal linewidth (bottom right) trace
$\sigma_{\rm{NT}}^2\propto r$ and $\sigma_{NT}^2\propto r^{1/2}$.  Recall
that $r_0^2\equiv9\sigma_c^2/4\pi G\rho_c$.
\label{fig1}}

\figcaption[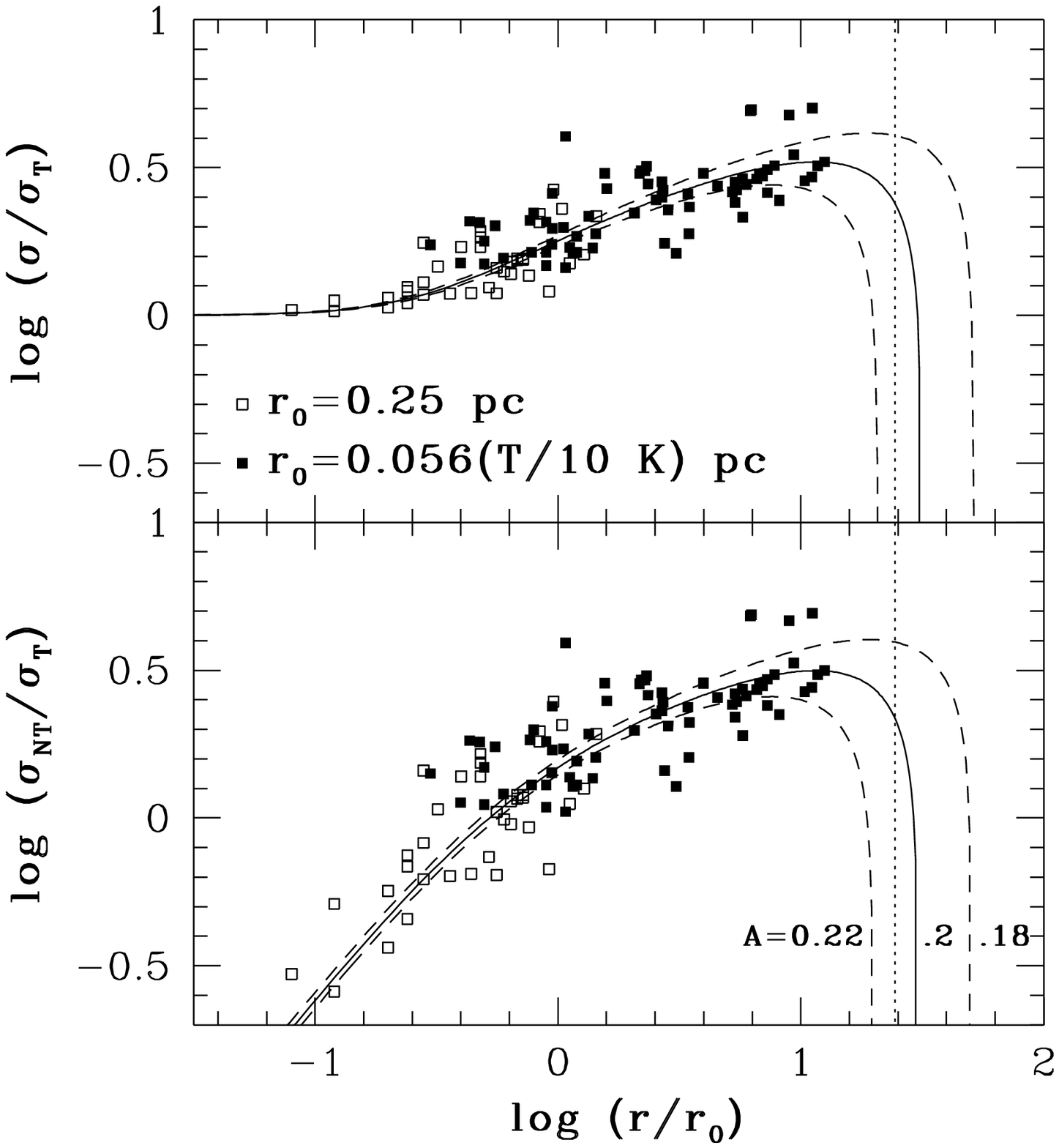]{Comparison of observed internal velocity-dispersion
profiles (total $\sigma$, thermal and nonthermal components $\sigma_{\rm{T}}$
and $\sigma_{\rm{NT}}$) with logotropic models, for 38 GMC cores. Open
squares correspond to low-mass cores (Fuller \& Myers 1992), and filled
symbols to high-mass cores (Caselli \& Myers 1995; we omit four HCO$^+$
measurements). Cores both with and without stars are represented here,
and each has been observed in three or more different molecular lines. The
model curves are for $A=0.20$ (best case; solid line), and $A=0.18$,
$A=0.22$ (dashed lines); a larger $A$ results in smaller maximum and
critical radii $r/r_0$.  The vertical line is at $\xi_{\rm{crit}}=24.37$ for
an $A=0.2$ logotrope.
\label{fig2}}

\clearpage

\plotone{fig0.eps}

\clearpage

\plotone{fig1.eps}

\clearpage

\plotone{fig2.eps}

\end{document}